# New and fast route to black $TiO_2$ based on hollow cathode $H_2$ plasma

A. Godoy Jr[#a], A.L.J. Pereira[†], M. C. Gomes[‡], R. S. Pessoa[#], D. M. G. Leite[#], G. Petraconi Filho[#], W. Miyakawa[§], and A S. da Silva Sobrinho[#b]

[#]Plasmas and Processes Laboratory - LPP, Technological Institute of Aeronautics - ITA – Praça Mal. Eduardo Gomes, 50, São José dos Campos/SP – Brazil
[†]Photonic Materials and Renewable Energy research group (MaFER), Grande Dourados Federal University – UFGD – Dourados/MS – Brazil
[‡] São Paulo Federal Institute – IFSP – Rod. Pres. Dutra, km 145, São José dos Campos/SP – Brazil
[§] Institute of Advanced Studies, Trevo Coronel Aviador José Alberto Albano do Amarante, 1, São José dos Campos/SP – Brazil

*Corresponding author e-mails:* godoyajr@gmail.com[a], argemiro@ita.br[b]

## Abstract

In this work, we demonstrate a new method to produce black $TiO_2$ from pristine anatase $TiO_2$ films. It consists on the immersion of $TiO_2$ films in a hollow cathode $H_2$ RF plasma for a few minutes, resulting in an efficient blackening of $TiO_2$. In this study, the pristine anatase $TiO_2$ films were grown by magnetron sputtering onto cover glass and c-Si substrates and then annealed at 450 °C for 2 h. Before and after the hollow cathode $H_2$ plasma treatment, the samples were characterized by profilometry, UV-Vis spectrophotometry, X-ray diffraction, Raman spectroscopy, X-ray photoelectron spectroscopy, field emission scanning electron and atomic force microscopies, and four-point probe measurements. The results show that the obtained black $TiO_2$ thin films have a significant light absorption on the whole solar spectrum, a very low sheet resistance, and also a relatively high surface area when compared to the pristine $TiO_2$. All these characteristics lead to an important improvement on their photocatalytic activity, as measured by the degradation rate of methylene blue under UV irradiation.

## Keywords

# 1  Introduction

As a result of the non-rational use of Earth's resources over the centuries, water, energy and food are now urgent problems of humanity to be solved, a fact that has encouraged many studies aimed at the synthesis and treatment of materials suitable for applications for this purpose. Specifically about water issue, semiconductor photocatalysis has shown great potential in sustainable technology to remove contaminants from liquids since 1972, when the photo-induced decomposition of water was discovered by Honda and Fujishima [1]. In this context, titanium dioxide ($TiO_2$) is one of the most studied semiconductor materials for this application due to its main characteristics: abundant, non-toxic, superior chemical stability, photocorrosion resistance, versatile functional material, and commercially inexpensive [2–4]. However, this material is also characterized by a wide band gap (~3.0-3.4 eV) limiting its photocatalytic activity only to the ultraviolet (UV) region [2], a small portion (∼5%) of the whole solar spectrum reaching the Earth's surface [5–8]. Therefore, different strategies, such as the insertion of metals or non-metal dopants in $TiO_2$-based materials, have been carried out to enhance the absorption of this material not only in the UV range but also over the visible spectrum [9,10,19–23,11–18]. However, this type of strategy is still the subject of much discussion since dopants can produce recombination centers that impair the efficiency of photoinduced processes [24]. In this scenario, Chen et al [2] synthesized black $TiO_2$ nanoparticles in 2011 showing a large optical absorption in the UV, visible and infrared region. The synthesis of this material required extreme process conditions such as high pressure of hydrogen gas (up to 20 bar) and long annealing treatment (up to 5 days).

Since then, many studies, as heat treatment at low and high pressure using different gases [2,4,33–42,25,43–47,26–32], high-energy particles bombardment (laser, electron) [48–50], chemical vapor deposition, plasma treatment [51–59], and others [60] have been focused to reduce the work pressure, due to the dangerous nature of $H_2$ under high pressure, or the use of other gases combined with hydrogen, as well as the reduction of process time. In view of this, the plasma process began to be applied on the $TiO_2$ blackening process in 2013 [52,54,61–63] with the propose to obtain this material with the expected characteristics in a short time by an efficient treatment [55,56,71–74,57,64–70]. Besides that, when carried out in vacuum environment, there are no issue in use $H_2$ gas, in addition to providing a process free of contamination.

The main characteristics expected after the black $TiO_2$ synthesis are disordered surface [52,75–78], presence of $Ti^{3+}$ and oxygen vacancies [4,30,33,70,79–83], Ti-H and O-H groups, and valence band edge. Due its interesting properties, black $TiO_2$ based materials can be applied in several areas, such as photocatalysis [31,48,89–91,61,62,77,84–88], hydrogen generation [92–95], solar desalination [96,97], dye sensitized solar cells [98,99], supercapacitors [100–102], batteries [45,54,94,95,103–108], and photothermal therapy [49,109,110].

Herein, we report a new and fast method for obtaining black $TiO_2$ based on $H_2$ RF plasma generated in a hollow cathode geometry. Hollow cathode plasmas are known to be several times denser [111] than a common planar discharge, thus leading to a more efficient hydrogen incorporation. In this scenario, the proposed experimental setup and process were very successful for blackening $TiO_2$ in a very short period of time. The films properties, as well as their performance as photocatalytic medium are then explored, confirming the good quality of the samples and the promising potential of the proposed technique in fabricating practical black $TiO_2$.

## 2 Experimental Method

### 2.1 Pristine $TiO_2$ thin film growth

Pure and transparent $TiO_2$ thin films were grown by DC magnetron sputtering (MS) onto cover glass and c-Si (100) substrates. Prior to deposition, the glass substrates were cleaned by sonication for 8 min with three solvents: ethyl alcohol, acetone, and isopropyl alcohol. The c-Si substrates were cleaned by a solution of sulfuric acid with hydrogen peroxide ($H_2SO_4/H_2O_2$, 4:1) for 10 min; after the c-Si were washed with deionized water and submitted to a solution of sulfuric acid with deionized water ($HF/H_2O$, 1:10) for 1 min and then washed with deionized water. Subsequently, the substrates were placed into the deposition chamber at a distance of 30 mm from the target.

Before each deposition, the chamber was evacuated using a diffusion + mechanical pump system that allows to reach a residual pressure of $5.0 \times 10^{-3}$ Pa. The sputtering process was performed using a 1.5-inch diameter titanium target (99.999%) in an $Ar+O_2$ (both 99.999%) atmosphere. The plasma was generated by a direct current (DC) power source (Advanced Energy model MDX 1K). All depositions were performed without external heating. However, at the end of the deposition, it was observed that the temperature of the substrates reached approx. 100 ºC. The main deposition parameters values are listed in **Table I**.

In order to assure the dominance of anatase phase, the produced films were annealed at 450 °C for 2 h in a tubular furnace at ambient pressure under room atmosphere. Here, the temperature rise was kept at 10 °C/min, while the cooling at approximately -2 °C/min.

## 2.2 Hollow cathode hydrogen plasma treatment

The hydrogen plasma process was performed in a capacitive plasma reactor (**Fig. 1a**) using a 13.56 MHz ENI model ACG-10B-01 radio frequency (RF) power source. The main improvement related to the experimental setup consists on the installation of a cylindrical hollow cathode (diameter = 30 mm, length = 120 mm), wherein the samples were placed. To avoid undesired contamination, the hollow cathode was entire built in titanium (99.99%).

The chamber was evacuated using a roots plus mechanical pump system (Edwards), reaching the residual pressure of $5.0 \times 10^{-2}$ Pa. The parameters used are summarized in **Table II**, and the samples were named based on the process time; e.g., "Black 15" corresponds to the sample treated with the hydrogen plasma for 15 min. The non-hydrogenated film was named as Pristine-$TiO_2$. As can be seen in **Fig. 1b**, after the hydrogen plasma treatment the films become black, even in the shortest treatment time of 15 min. Using an optical pyrometer (Raytek, Raynger 3I), it was inferred that the interior of the hollow cathode wall reaches a temperature of about 256 $^o$C after 8 min and maintain this temperature up to the end of the treatment. Is important to highlight the importance of the hollow cathode in producing black $TiO_2$: the same treatment performed using the parameters presented in **Table II** for 240 min without the hollow cathode did not present significant alteration in the properties of the $TiO_2$ film (**Fig. 1c**). It happens because the plasma processes that occurs inside a hollow cathode are more intense [111], increasing the efficiency in the incorporation of hydrogen in the films. In synthesis, this is a consequence of the high energy electrons oscillating motion between repelling potentials of the walls in the cathode [111,112] and due to the secondary electrons emitted from the cathode surface wall - originated mainly from positive ions impacting the cathode surface - be accelerated by the electric field across the plasma [111]. These electrons can suffer several collisions with the gas along their paths enhancing ionization and consequently the plasma density.

## 2.3 Material characterization

The thickness of the films was measured by a mechanical profilometer (KLA Tencor P-7) and by spectrophotometric ellipsometry using a HORIBA UVISEL II. The surface morphologies were analyzed by AFM (Shimadzu SPM 9500J3) and FEG-SEM (Mira 3 Tescan). The film surface area was obtained by the Data Processing Software (SPM-9500 Series, Version 2.4, Shimadzu).

Structural characterization of the films was carried out by X-ray Diffraction (XRD) performed with a PANalytical Empyrean diffractometer using Cu $K_\alpha$ (1.5406 Å for $K_{\alpha 1}$) as the incident radiation source. The scanning speed was 3.2°/min in the range of 20−80°. The lattice parameters were determined by the HighScore Software [113].

Vibrational characterization was carried out by Raman scattering (RS) measurements performed with a Horiba Evolution microspectrometer, equipped with a thermoelectrically cooled multichannel charge-coupled device detector which allows a spectral resolution better than 1 cm$^{-1}$. The signal was collected in backscattering geometry using a 100x objective. The excitation was done by a 532 nm laser (power < 10 mW). Phonons modes were then analyzed by fitting Raman peaks with a Voigt profile fixing the Gaussian linewidth (1.6 cm$^{-1}$) to the experimental setup resolution.

In order to analyze the surface composition and evaluate the influence of the plasma treatment in the $TiO_2$ thin films, X-ray photoelectron spectroscopy (XPS) measurements were performed in a Kratos Axis Ultra spectrometer using an Al $K_\alpha$ (hv = 1486.69 eV) excitation source operating at 120W.

The sheet resistivity of the films was characterized by four-point probe measurements performed in a JADEL test unit, model RM3000. The measurements were performed at ambient pressure using 100 μA in three different points of the sample and the mean values were considered.

Finally, the optical characterizations were performed using UV-Vis-NIR transmittance measurements. The transmittance measurements were carried out in the 190-2500 nm range at normal incidence in a Jasco V-570 spectrophotometer equipped with 150 mm integrating sphere apparatus. The optical band gap was analyzed by Talc plot method. This method consists in the extrapolation of a straight-line portion of the $(\alpha h\nu)^{\frac{1}{2}}$ versus photon energy ($h\nu$) of incident light. The optical absorption coefficient $\alpha$ was calculated by the follow equation:

$$\alpha = -\frac{1}{d} \ln\left(\frac{T}{1-R}\right) \quad (1)$$

where $T$ is the transmittance, $R$ the reflectance and $d$ the film thickness [114–116].

## 2.4 Photocatalytic activity measurement

The photocatalytic activity was evaluated by detecting the degradation rate of methylene blue (MB) using a homemade reactor with six UV lamps of 15W (HNS G13 - G15T8/OF) used as a light source with an air-cooling system. Samples grown on glass, with an effective area of 20 mm × 20 mm, were placed in 6 mL of 10 mg/L MB solution. The solution with the sample was stirred for 1 h in the dark and was cooled using a water circulating jacket maintaining the solution at room temperature. After this period, the lamps were turned on and the mixture was maintained stirred constantly under the UV light at a distance of 70 mm. The degradation of MB solution was measured by the UV−vis spectrophotometer (Thermo Scientific Evolution 220) every 10 min.

# 3 Results and discussion

As the main result, the proposed methodology showed to be an efficient route to produce black $TiO_2$ thin films with interesting properties for a wide range of applications, including solar cell, photocatalysis and others. The films properties results are separated according to the physical and chemical characteristics evaluated. At the end, the results concerning the performance of the black $TiO_2$ as photocatalytic medium are demonstrated and discussed.

### 3.1 Black $TiO_2$ morphology

In **Fig. 2**, field emission gun scanning electron microscopy (FEG-SEM) imagens are displayed. Pristine $TiO_2$ micrograph (**Fig. 2a**) shows a surface formed by small round particles, homogeneously distributed over the surface. Comparing with pristine $TiO_2$, films surface after hydrogen plasma treatment seemed not to be significantly altered, except for some slight cracks and a "kneaded" aspect, which was more intense as the time of treatment increased. AFM images (**Fig. 3**) presented good agreement with FEG-SEM observations. Pristine $TiO_2$ surface is formed by the superposition of round particles approximately 100 nm long (**Fig. 3a**). **Fig. 3b** (15 min of treatment) seemed to be formed of densely packed clusters of smaller round particles (~ 40-50 nm long). In **Fig. 3c**, **d**, and **e**, clusters are becoming proportionally larger and particles, proportionally shrunken. Considering that the total effective surface area is a demand for catalytic processes, **Table III** shows the obtained values for this parameter using the AFM facility. All treated samples presented larger surface area than the untreated one, having the Black 15 sample the greater increase (~24 % of the surface area) and the other, about 10% of increase. This result can be a consequence of the etch of the small grains due the removal of the oxygen by the H+ radicals [117,118]. In fact, profilometry measurements after the plasma treatment indicated a decrease of ~0.6 nm/min in the films thickness. Hence, it can be concluded that for treatment times greater than 15 min, the hydrogen ions have possibly created OH volatile species on the films surface, causing a surface planning.

## 3.2 Black TiO$_2$ structural properties

In order to evaluate the effects of the hydrogen treatment in the crystalline structure, XRD measurements were performed in pristine and black TiO$_2$ (**Fig. 4**). Only tetragonal anatase phase related peaks [119–121] are observed for all samples. Their FWHM indicate a polycrystalline structure. The main diffraction peak, related to the (101) plane, of the black TiO$_2$ samples presents a small shift (~0.1°) to lower angles when compared to the pristine TiO$_2$, which results in a small increase on the lattice parameters: $a$ = 3.79 Å and $c$ = 9.49 Å for the former and $a$ = 3.77 Å and $c$ = 9.46 Å for the latter. Furthermore, black TiO$_2$ films exhibited an additional broadening of the anatase (101) peak besides the decrease of its intensity as the hollow cathode treatment time increased. These changes indicate an increase of the disorder in the hydrogenated black TiO$_2$, which probably occurs due the generation of Ti$^{3+}$ and Ti$^{2+}$ species [122,123], as will be discussed in more detail below.

In order to complement the structural characterization, the samples were also characterized by Raman scattering (**Fig. 5**). The Raman spectra displays five vibrational modes observed around 139.5 cm$^{-1}$, 199.0 cm$^{-1}$, 390.5 cm$^{-1}$, 513.3 cm$^{-1}$, and 632.6 cm$^{-1}$ for all samples. These peaks correspond to the anatase active vibrations mode [124–126], reinforcing the XRD results. In addition, the temperature increase produced by the plasma process was not enough to produce some rutile phase fraction. The main modifications occurred after plasma treatment were: (i) a blue-shift and (ii) a broadening of the FWHM of the most intense E$_g$ mode (see **Fig. 5b**). These phenomena were previous described and attributed to the formation of oxygen vacancies (V$_o$) on the TiO$_2$ surface, indicating a decrease in the original symmetry of TiO$_2$ after the hydrogenation process [75,77,127,128].

In order to investigate the surface chemical composition of the samples, XPS measurements were taken. **Fig. 6** shows the high resolution XPS spectrum of the Ti 2p$_{3/2}$ of pristine and black TiO$_2$ samples. All the spectra were corrected for the C1s peak at 284.6 eV for determination of binding energy (BE) values of different elements. We can highlight that no metallic Ti was observed both before and after the treatment in hydrogen plasma. Is possible to observe that the Ti 2p$_{3/2}$ spectrum of the pristine TiO$_2$ films (**Fig. 4a**) presents only a peak at 459.69 eV which is related to Ti with oxidation state 4+ [64,71,94,129–134]. After plasma treatment, the high resolution XPS spectra show a shift of ~0.7 eV at low energies and we could also detect two new peaks (see **Fig. 4b-e**) that can be attributed mainly to 2p$_{3/2}$ Ti$^{3+}$ and Ti$^{2+}$ species [70,135]. As can be seen in **Table IV**, the sample treated for 15 min presents almost the same concentration of Ti$^{3+}$ and Ti$^{2+}$ species. However, as the treatment time increases, the concentration of Ti$^{3+}$ and Ti$^{2+}$ tends to increase while the concentration of Ti$^{4+}$ tends to decrease.

High resolution XPS spectra of the O1s core level for pristine and black $TiO_2$ films are presented in **Fig. 7** and its BE positions are tabulated in **Table V**. The O1s spectrum of all samples can be deconvoluted into two peaks: the first located around 532.1 eV and the second centered around 530.8 eV. As can be seen in **Table V**, after the treatment in hydrogen plasma, there is a significant increase in the area of the peak centered at 532.1 eV and a decrease of the peak at 530.8 eV. The peak at 530.8 eV is mainly related to the $O^{2-}$ anion in $TiO_2$ [136] and the peak at 532.1 eV is generally associated with hydroxyl groups [37,38] or $V_o$ in titanium oxide ($TiO_x$) [40]. In the case of the pure $TiO_2$ film, both options are possible. However, since the annealing at 450 $^o$C was performed at ambient pressure, the creation of $V_o$ should not be high [135]. This leads us to believe that the peak at 532.1 eV in the pristine $TiO_2$ film is related mainly to the hydroxyl groups adsorbed on the surface of the film. The increase of this peak intensity with the hydrogen plasma treatment may be related mainly to the creation of $TiO_x$ defect. As can be observed in **Fig. 8a**, after exposing the film to a hydrogen plasma for 15 min, the Ti-O ratio decreases ~1.5% relative to the O-H and/or $TiO_x$ species. For longer treatment times, there is no significant variation of this ratio, indicating that 15 min are enough to produce significant variations on the chemical composition of the film surface. In **Fig. 8b** is possible to observe that, after the hydrogen plasma treatment, the % area of the 532.1 eV peak increases ~ 20% while the concentration of $Ti^{2+/+3}$ increases ~ 25%. This is a strong indication that most of the reaction between $H^+$ and $TiO_2$ film surface results in the desorption of O from the film surface, probably in the form of water molecules, creating $V_o$ [135].

Hannula et al. pointed out that the $Ti^{2+}$ is only observed after an annealing above 400 $^o$C in an ultra-vacuum chamber or after 5 min at 300 $^o$C in a hydrogen atmosphere [135]. In this study, even the sample temperature not exceeding 250 $^o$C, the same concentration of $Ti^{2+}$ and $Ti^{3+}$ was observed for the 15 min treatment. This behavior may be directly related to the high energy transferred to the thin film during the hollow cathode plasma process that allows the direct reduction of $Ti^{4+}$ to $Ti^{2+}$ or $Ti^{3+}$ at the same rate. Thus, the Ti reduction process by means of the hydrogen plasma treatment that was performed herein can be summarized according to the scheme presented in **Eq. 2** and **Eq. 3**.

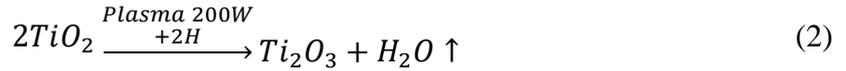

$$2TiO_2 \xrightarrow[+2H]{Plasma\ 200W} Ti_2O_3 + H_2O \uparrow \qquad (2)$$

or

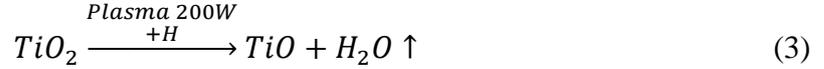

$$TiO_2 \xrightarrow[+H]{Plasma\ 200W} TiO + H_2O \uparrow \qquad (3)$$

As it was already discussed, the hollow cathode plasma treatment of $TiO_2$ films results in a reduction of the $Ti^{+4}$. The removal of one oxygen leaves an excess of electrons that occupy energy levels close to the conduction band produced by the $V_o$ [70], increasing the electrical conductivity. In fact, through sheet resistivity measurements (**Table III**) we observed a significant decrease in resistivity after 15 min of plasma treatment. As the treatment time increases, the sheet resistivity decreased at a rate of -21.5 $\Omega/\square \cdot min^{-1}$ approximately. Considering that the measuring range of the equipment is $10M\Omega/\square$, the resistivity reduction was at least 99.8 %.

### 3.3 Optical and electric properties

Considering the results presented so far, it is clear that the treatment of $TiO_2$ films in hydrogen plasma using a hollow cathode is quite efficient in the production of black $TiO_2$. Only 15 min of this process is enough for a significant creation of $V_o$. The darkening of the samples is an indicative of an increase in the absorption of visible light. As photoinduced phenomena in $TiO_2$ are not very efficient in the visible range of the electromagnetic spectra, this increase is welcome in applications such as photocatalysis and solar cells. In order to study the effects of hydrogen plasma treatment in the optical properties of $TiO_2$ films, we performed UV-vis transmittance measurements (see **Fig. 9a**). The spectra confirm that pristine $TiO_2$ absorbs light only below 400 nm (3.1 eV), i.e., in the UV region. On the other hand, the hydrogenated films, besides the strong absorption in the UV region, also present a significant absorption, proportional to the process time, in the region of the visible and the near IR (**Fig. 9a**).

**Figures 9b-f** show the Tauc plots for all investigated samples. It is observed that bandgap energy of all films is around 3.3 eV, indicating that the hydrogen plasma treatment produces no significant variation in the distance between the valence band (VB) and the conduction band (CB). However, the absorption in the visible range can be attributed to the induced $Ti^{3+}$ and $Ti^{2+}$ defects [30,33,52,75,119,137–140], which is responsible to a n-type energy level in the bandgap of black $TiO_2$ films [75,129]. In addition, the presence of $V_o$ is responsible for increase the disorder of the material, which may increase the Tail States (TS) at the edges of the valence and conduction band [75].

In order to investigate in more detail, the changes produced in VB, we performed High-resolution Ultraviolet Photoelectron Spectroscopy (UPS - **Fig. 10**) which provides a spectrum proportional to the occupied density of state (DOS) in the top of VB and was used to investigate the electronic structures of the pristine and black $TiO_2$ films. Extrapolating to zero the linear region of the lower energy side at the top of the VB (**Fig. 10**) we obtain a binding energy of ~3.38 eV to the pure $TiO_2$ and ~3.25 eV to all black $TiO_2$, which made the values of intrinsic band-gap of each sample only have slightly decrease with the hydrogen plasma treatment. These results are in good agreement with the values of bandgap obtained by the Tauc's model (**Fig. 9**). In addition, UPS measurements allow the estimation of TS size related to the disorder produced by $V_o$ [91]. Despite not having undergone the treatment in hydrogen plasma, pristine $TiO_2$ have a TS of 0.29 eV (assigned as $\Delta$ in **Fig. 10**). The intrinsic defects of films grown by sputtering, like lattice defects, residual stress and grain contour, can be responsible for the presence of TS in the pristine film [141,142]. As the time of treatment in hydrogen plasma increases, the size of the tail states ($\Delta$) also tends to increase in a rate of $6.8 \times 10^{-3}$ eV.min$^{-1}$. The increase of TS can be directly related to the formation of $Ti^{3+/2+}$ discussed earlier.

As it can also be observed in **Fig. 10**, the top of VB of all films was fitted using five energy states. The level that we call $\gamma$, at 10.3 eV (in pristine $TiO_2$), present an additional contribution of 0.11 %.min$^{-1}$ of the plasma treatment. This peak can be associated to OH impurities adsorbed in the surface of the films [135,143,144]. Another possibility is a residual Si impurity that segregate from the substrate to the surface [135,145]. However, this last possibility is unlikely, since the Si segregation was only observed in materials submitted to temperatures higher than 800 °C [135]. Hannula et al. reported that this peak disappears after an annealing at 400 °C in vacuum or after 1 min of a treatment at 300 °C in hydrogen atmosphere [135]. In the present case, this contribution increases with the treatment time, indicating that the energies involved in the process are not enough to eliminate OH impurities from the surface of the films. At the same time, the so called O2p nonbonding contribution, related to nonbonding O2p orbitals [135], tends to decrease at a rate of -0.14 %.min$^{-1}$ of treatment. This is in agreement with our XPS measurements that indicate a removal of O atoms from the film surface during treatment in hydrogen plasma (see previous section). These results then indicate that two phenomena are happening at the same time and practically at the same rate: i) volatization of the oxygen; and ii) incorporation of OH species in the surface of the films. The fact that the treatment is promoted by a $H_2$ plasma generated inside a hollow cathode geometry may explain this phenomenon. When the ions $H^+$ react with O of the film surface, it volatizes as H$_2$O. However, as it is relatively confined inside the cathode, the water molecule can re-ionize, forming HO$^-$ species that are incorporate on the film surface.

The peak at 7.7 eV and 5.7 eV (in pristine TiO$_2$) are the σ- and π-type molecular orbitals of TiO$_2$, respectively [135]. The σ-type is mainly related to the O2p$_z$ and the π-type to the O2p$_x$ and O2p$_z$ [146,147]. As can be observed in **Fig. 10**, the concentration of σ-type orbitals in the pristine TiO$_2$ is majoritarian in the top of the VB, a common behavior in pristine TiO$_2$ [135]. However, as the time of the treatment in hydrogen plasma increases, the contribution of the π-type orbitals increases while the σ-type decreases at the same rate (0.67%/min). In addition, it is possible to observe that the TS are mainly caused by the increase of the states of the p-type bonds.

The occupied band centered at 0.8 eV, just below the Fermi level, is commonly attributed to mid-gap states produced by dopants, Ti$^{2+/3+}$ interstitial, or V$_o$ [135]. In the pristine TiO$_2$ sample, the presence of these states contributes only with 0.6% of the states present at the top of the VB. The presence of this peak in pristine TiO$_2$ film agrees with the O1s states discussed in **Fig. 7a** and **Fig. 8** and can be related to the intrinsic TiO$_2$ defects. After treatment in hydrogen plasma for 15 min, this contribution significantly increases up to 2.9%. The position of this peak remains practically constant at 0.8 eV independent of the treatment time and is in agreement with the value theoretically obtained for Ti$^{3+}$ (0.8 eV), V$_o$ (0.7 eV) and Ti$^{3+}$ -OH (0.4 eV) [148], thus indicating the creation/increase of the Ti$^{2+/3+}$ states observed in **Fig. 6**. These states, together with the tail states, provide a shorter-energy transitions to CB than the direct band-gap transitions, explaining the sub-gap absorption in visible and NIR (**Fig. 9**).

### 3.4 Photocatalytic activity

The photocatalytic activity of the pristine and black TiO$_2$ thin films was studied by monitoring the changes in the optical absorption of MB under UV irradiation. Before turning on the UV light, the film was held in the solution for 2h to achieve the adsorption/desorption equilibrium. After this time, the first absorbance measurement was made, which we called C$_0$. Then, the UV light was switched on and a measurement was performed every 10 min. In **Fig 11** it is possible to observe a linear behavior for all the samples, which allows use the following equation:

$$\ln\frac{C_0}{C} = kt \qquad (4)$$

where $C_0$ is the initial MB concentration, $C$ is the concentration after a determined period of UV light irradiation, $t$ represents the irradiation time, and $k$ apparent rate constant (min$^{-1}$) (**Fig. 11b**) [123,149]. The constant $k$ is known to be independent of the dye concentration or temperature, considering that the photodegradation is a function that depends only of the irradiation flux and light source spectrum [149,150].

Analyzing **Fig. 11**, it is important to highlight that the pure solution did not exhibit any considerable photocatalytic activity under UV exposure. We also can observe that all the black $TiO_2$ films present improved photocatalytic activity in comparison with pristine $TiO_2$ film. This result may be a direct consequence of the changes caused by the presence of $Ti^{2+/3+}$, $V_o$ and the increase of hydroxyl radicals (OH) on the film surface, as evidenced by the XPS and UPS results. The OH groups is the mainly oxidative species in photocatalysis reactions, once these groups have a great performance on the decomposition of organic molecules [19,119,123,151–154]. Another factor that contributes to the increase of the photocatalytic activity after plasma treatment can be attributed to the increase of the surface area of the black films in relation to the pristine ones, as shown in the SEM and AFM results. It has been reported that the surface area is a determining factor in increasing dye degradation [19]. Besides, the 15 min and 30 min black $TiO_2$ films presented the best performance on the MB rate degradation compared with 45 min and 60 min black $TiO_2$ samples. According to **Table V**, considering the OH, we can observe that there is no significant difference between the films treated with hydrogen plasma, independent of the treatment time. Therefore, the hydroxyl groups concentration is not the main factor responsible for the best performance of the 15 min sample compared to the other films. However, as discussed above, surface area is a determinant parameter on the photocatalytic activity performance, and was evidenced in SEM and AFM results that the highest surface area value was obtained for the film treated during only 15 min. Therefore, because this highest value of surface area that the 15 min sample performed the highest MB photodegradation rate. Finally, the increase of the number of oxygen vacancies and $Ti^{2+/3+}$ species on the black $TiO_2$ films also contributed to the photodegradation, and this are responsible to the optical and electronic properties enhanced [4,30,33,70,79–83]. In view of this, it is expected a best photocatalytic activity when these black $TiO_2$ films will be applied on visible light photocatalysis.

# 4 Conclusion

A promising method based on hollow cathode $H_2$ plasma was proposed and proven to be a fast and efficient way for blackening $TiO_2$ films. The hydrogen plasma treatment leads to a significant change in the microstructural, optical, morphological, and electronic properties. These properties modifications are mainly related to the increase of oxygen vacancies, $Ti^{3+}$ species, Ti-H, and OH- species on the black $TiO_2$ film lattice. The creation of mid gap states between the valance and conduction bands besides the creation of tail states on the valance band were responsible for the important sub-gap absorption extended from UV to NIR over all the visible range. In addition to the optic-electronic modifications, an important increase on the specific surface area was also obtained as a direct result from the hollow cathode plasma treatment. All these improvements reflected positivity when the black $TiO_2$ films were applied on photocatalysis experiments under UV light. In general, it was possible to conclude that the shortest treatment time (15 min) was enough to enhance the microstructural properties. This confirms the hollow electrode $H_2$ plasma with an efficient and rapid method to blackening $TiO_2$ thin films. Moreover, when the treatment time was superior to 15 min, the black $TiO_2$ films presented a specific surface area decrease, which was averse to its photodegradation performance. Finally, it is expected, based on the properties acquired, a best photocatalytic activity when these black $TiO_2$ thin films will be applied on visible light photocatalysis.

## Acknowledgments

The authors acknowledge the financial support of CAPES, CNPq and FAPESP (Grant 2018-6/159754, 2018-5/307199, 2012-9/201050, 2010-8/555.686, 2012-8/470820 and 2015/06241-5). The authors also thank the Associated Laboratory of Sensors and Materials – LABAS of National Institute of Space Research - INPE for the XPS and FEG-SEM analysis.

## References

[1] A. Fujishima, K. Honda, Electrochemical photolysis of water at a semiconductor electrode, Nature. 238 (1972) 37–38. doi:10.1038/238037a0.

[2] C. Sun, Y. Jia, X.H. Yang, H.G. Yang, X. Yao, G.Q. Lu, A. Selloni, S.C. Smith, Hydrogen incorporation and storage in well-defined nanocrystals of anatase titanium dioxide, J. Phys. Chem. C. (2011). doi:10.1021/jp210472p.

[3] T. Bak, J. Nowotny, M. Rekas, C.C. Sorrell, Photo-electrochemical hydrogen generation from water using solar energy, Int. J. Hydrogen Energy. (2002). doi:http://dx.doi.org/10.1016/S0360-3199(02)00022-8.

[4] G. Wang, H. Wang, Y. Ling, Y. Tang, X. Yang, R.C. Fitzmorris, C. Wang, J.Z. Zhang, Y. Li, Hydrogen-Treated $TiO_2$ Nanowire Arrays for Photoelectrochemical Water Splitting, Nano Lett. 11 (2011) 3026–3033. doi:10.1021/nl201766h.

[5] M.A. Mansoor, M. Mazhar, A. Pandikumar, H. Khaledi, H. Nay Ming, Z. Arifin, Photoelectrocatalytic activity of $Mn_2O_3$-$TiO_2$ composite thin films engendered from a trinuclear molecular complex, Int. J. Hydrogen Energy. 41 (2016) 9267–9275. doi:10.1016/j.ijhydene.2016.04.121.

[6] C. Dette, M.A. Pérez-Osorio, C.S. Kley, P. Punke, C.E. Patrick, P. Jacobson, F. Giustino,

S.J. Jung, K. Kern, TiO$_2$ anatase with a bandgap in the visible region, Nano Lett. 14 (2014) 6533–6538. doi:10.1021/nl503131s.

[7] X. Chen, A. Selloni, Introduction: Titanium dioxide (TiO$_2$) nanomaterials, Chem. Rev. 114 (2014) 9281–9282. doi:10.1021/cr500422r.

[8] Y. Liu, L. Tian, X. Tan, X. Li, X. Chen, Synthesis, properties, and applications of black titanium dioxide nanomaterials, Sci. Bull. 62 (2017) 431–441. doi:10.1016/j.scib.2017.01.034.

[9] D.A. Duarte, M. Massi, A.S. Da Silva Sobrinho, Development of dye-sensitized solar cells with sputtered N-Doped TiO$_2$ thin films: From modeling the growth mechanism of the films to fabrication of the solar cells, Int. J. Photoenergy. (2014). doi:10.1155/2014/839757.

[10] A. Sarkar, A. Shchukarev, A.R. Leino, K. Kordas, J.P. Mikkola, P.O. Petrov, E.S. Tuchina, A.P. Popov, M.E. Darvin, M.C. Meinke, J. Lademann, V. V. Tuchin, Photocatalytic activity of TiO$_2$ nanoparticles: Effect of thermal annealing under various gaseous atmospheres, Nanotechnology. (2012). doi:10.1088/0957-4484/23/47/475711.

[11] R. Marschall, L. Wang, Non-metal doping of transition metal oxides for visible-light photocatalysis, Catal. Today. (2014). doi:10.1016/j.cattod.2013.10.088.

[12] J. Mu, B. Chen, M. Zhang, Z. Guo, P. Zhang, Z. Zhang, Y. Sun, C. Shao, Y. Liu, Enhancement of the visible-light photocatalytic activity of In$_2$O$_3$-TiO$_2$ nanofiber heteroarchitectures, ACS Appl. Mater. Interfaces. (2012). doi:10.1021/am201499r.

[13] F. Zhou, H. Song, H. Wang, S. Komarneni, C. Yan, N-doped TiO$_2$/sepiolite nanocomposites with enhanced visible-light catalysis: Role of N precursors, Appl. Clay Sci. 166 (2018) 9–17. doi:10.1016/j.clay.2018.08.025.

[14] R.S. Pessoa, M.A. Fraga, L. V. Santos, M. Massi, H.S. Maciel, Nanostructured thin films based on TiO$_2$ and/or SiC for use in photoelectrochemical cells: A review of the material characteristics, synthesis and recent applications, Mater. Sci. Semicond. Process. 29 (2015) 56–68. doi:10.1016/j.mssp.2014.05.053.

[15] T. Zhang, B. Yu, D. Wang, F. Zhou, Molybdenum-doped and anatase/rutile mixed-phase TiO$_2$ nanotube photoelectrode for high photoelectrochemical performance, J. Power Sources. (2015). doi:10.1016/j.jpowsour.2015.02.017.

[16] D.A. Duarte, M. Massi, A.S.D.S. Sobrinho, Fabrication of dye-sensitized solar cells with N-doped TiO$_2$ thin films: Effect of the nitrogen doping on the photoexcitation processes and generation of electron acceptor states, in: Conf. Rec. IEEE Photovolt. Spec. Conf., 2013. doi:10.1109/PVSC.2013.6745030.

[17] M. Mollavali, C. Falamaki, S. Rohani, Preparation of multiple-doped TiO$_2$ nanotube arrays with nitrogen, carbon and nickel with enhanced visible light photoelectrochemical activity via single-step anodization, Int. J. Hydrogen Energy. 40 (2015) 12239–12252. doi:10.1016/j.ijhydene.2015.07.069.

[18] Y.M. Pan, W. Zhang, Z.F. Hu, Z.Y. Feng, L. Ma, D. ping Xiong, P.J. Hu, Y.H. Wang, H. yi Wu, L. Luo, Synthesis of Ti$^{4+}$-doped ZnWO4 phosphors for enhancing photocatalytic activity, J. Lumin. 206 (2019) 267–272. doi:10.1016/j.jlumin.2018.10.054.

[19] G. Wang, J. Zheng, H. Bi, S. Wang, J. Wang, J. Sun, Y. Guo, C. Wang, Ti$^{3+}$ self-doping in bulk of rutile TiO$_2$ for enhanced photocatalysis, Scr. Mater. 162 (2019) 28–32.

doi:S1359646218306444.

[20] D.A. Duarte, J.C. Sagás, A.S. Da Silva Sobrinho, M. Massi, Modeling the reactive sputter deposition of N-doped $TiO_2$ for application in dye-sensitized solar cells: Effect of the $O_2$ flow rate on the substitutional N concentration, in: Appl. Surf. Sci., 2013. doi:10.1016/j.apsusc.2012.10.048.

[21] Y. Chen, Q. Wu, L. Liu, J. Wang, Y. Song, The fabrication of self-floating $Ti^{3+}$/N co-doped $TiO_2$/diatomite granule catalyst with enhanced photocatalytic performance under visible light irradiation, Appl. Surf. Sci. 467–468 (2019) 514–525. doi:10.1016/j.apsusc.2018.10.146.

[22] Z. Zheng, B. Huang, X. Qin, X. Zhang, Y. Dai, M.H. Whangbo, Facile in situ synthesis of visible-light plasmonic photocatalysts $M@TiO_2$(M = Au, Pt, Ag) and evaluation of their photocatalytic oxidation of benzene to phenol, J. Mater. Chem. (2011). doi:10.1039/c1jm10983a.

[23] D.A. Duarte, M. Massi, J.C. Sagás, A.S. Da Silva Sobrinho, D.R. Irala, L.C. Fontana, Hysteresis-free deposition of $TiO_xN_y$ thin films: Effect of the reactive gas mixture and oxidation of the TiN layers on process control, Vacuum. (2014). doi:10.1016/j.vacuum.2013.08.014.

[24] A. Zaleska, Doped-$TiO_2$: A Review, Recent Patents Eng. (2008). doi:10.2174/187221208786306289.

[25] L. Liu, P.Y. Yu, X. Chen, S.S. Mao, D.Z. Shen, Hydrogenation and disorder in engineered black $TiO_2$, Phys. Rev. Lett. 111 (2013) 1–5. doi:10.1103/PhysRevLett.111.065505.

[26] X. Chen, L. Liu, F. Huang, Black titanium dioxide ($TiO_2$) nanomaterials, Chem. Soc. Rev. 44 (2015) 1861–1885. doi:10.1039/c4cs00330f.

[27] Y. Ma, X. Wang, Y. Jia, X. Chen, H. Han, C. Li, Titanium dioxide-based nanomaterials for photocatalytic fuel generations, Chem. Rev. 114 (2014) 9987–10043. doi:10.1021/cr500008u.

[28] M. Kapilashrami, Y. Zhang, Y.S. Liu, A. Hagfeldt, J. Guo, Probing the optical property and electronic structure of $TiO_2$ nanomaterials for renewable energy applications, Chem. Rev. (2014). doi:10.1021/cr5000893.

[29] M. Dahl, Y. Liu, Y. Yin, Composite titanium dioxide nanomaterials, Chem. Rev. 114 (2014) 9853–9889. doi:10.1021/cr400634p.

[30] X. Jiang, Y. Zhang, J. Jiang, Y. Rong, Y. Wang, Y. Wu, C. Pan, Characterization of oxygen vacancy associates within hydrogenated $TiO_2$: A positron annihilation study, J. Phys. Chem. C. 116 (2012) 22619–22624. doi:10.1021/jp307573c.

[31] T. Leshuk, R. Parviz, P. Everett, H. Krishnakumar, R.A. Varin, F. Gu, Photocatalytic activity of hydrogenated $TiO_2$, ACS Appl. Mater. Interfaces. (2013). doi:10.1021/am302903n.

[32] A. Sinhamahapatra, J.P. Jeon, J.S. Yu, A new approach to prepare highly active and stable black titania for visible light-assisted hydrogen production, Energy Environ. Sci. 8 (2015) 3539–3544. doi:10.1039/c5ee02443a.

[33] A. Naldoni, M. Allieta, S. Santangelo, M. Marelli, F. Fabbri, S. Cappelli, C.L. Bianchi, R. Psaro, V. Dal Santo, Effect of nature and location of defects on bandgap narrowing in

black TiO$_2$ nanoparticles, J. Am. Chem. Soc. 134 (2012) 7600–7603. doi:10.1021/ja3012676.

[34] K. Zhang, W. Zhou, X. Zhang, Y. Qu, L. Wang, W. Hu, K. Pan, M. Li, Y. Xie, B. Jiang, G. Tian, Large-scale synthesis of stable mesoporous black TiO$_2$ nanosheets for efficient solar-driven photocatalytic hydrogen evolution: Via an earth-abundant low-cost biotemplate, RSC Adv. 6 (2016) 50506–50512. doi:10.1039/c6ra06751d.

[35] N. Liu, C. Schneider, D. Freitag, M. Hartmann, U. Venkatesan, J. Müller, E. Spiecker, P. Schmuki, Black TiO$_2$ nanotubes: Cocatalyst-free open-circuit hydrogen generation, Nano Lett. (2014). doi:10.1021/nl500710j.

[36] T. Leshuk, S. Linley, F. Gu, Hydrogenation processing of TiO$_2$ nanoparticles, Can. J. Chem. Eng. 91 (2013) 799–807. doi:10.1002/cjce.21745.

[37] Z. Lu, C.T. Yip, L. Wang, H. Huang, L. Zhou, Hydrogenated TiO$_2$ nanotube arrays as high-rate anodes for lithium-ion microbatteries, Chempluschem. (2012). doi:10.1002/cplu.201200104.

[38] M. Ge, C. Cao, J. Huang, S. Li, Z. Chen, K.Q. Zhang, S.S. Al-Deyab, Y. Lai, A review of one-dimensional TiO$_2$ nanostructured materials for environmental and energy applications, J. Mater. Chem. A. (2016). doi:10.1039/c5ta09323f.

[39] S. Wei, R. Wu, X. Xu, J. Jian, H. Wang, Y. Sun, One-step synthetic approach for core-shelled black anatase titania with high visible light photocatalytic performance, Chem. Eng. J. 299 (2016) 120–125. doi:10.1016/j.cej.2016.04.067.

[40] Y. Zhu, D. Liu, M. Meng, H$_2$ spillover enhanced hydrogenation capability of TiO$_2$ used for photocatalytic splitting of water: A traditional phenomenon for new applications, Chem. Commun. 50 (2014) 6049–6051. doi:10.1039/c4cc01667j.

[41] M.C. Wu, I.C. Chang, K.C. Hsiao, W.K. Huang, Highly visible-light absorbing black TiO$_2$ nanocrystals synthesized by sol-gel method and subsequent heat treatment in low partial pressure H$_2$, J. Taiwan Inst. Chem. Eng. 63 (2016) 430–435. doi:10.1016/j.jtice.2016.02.026.

[42] L. Han, Z. Ma, Z. Luo, G. Liu, J. Ma, X. An, Enhanced visible light and photocatalytic performance of TiO$_2$ nanotubes by hydrogenation at lower temperature, RSC Adv. (2016). doi:10.1039/c5ra11616c.

[43] X. Liu, Z. Xing, H. Zhang, W. Wang, Y. Zhang, Z. Li, X. Wu, X. Yu, W. Zhou, Fabrication of 3D Mesoporous Black TiO$_2$/MoS$_2$/TiO$_2$ Nanosheets for Visible-Light-Driven Photocatalysis, ChemSusChem. 9 (2016) 1118–1124. doi:10.1002/cssc.201600170.

[44] L.R. Grabstanowicz, S. Gao, T. Li, R.M. Rickard, T. Rajh, D.J. Liu, T. Xu, Facile oxidative conversion of TiH$_2$ to high-concentration Ti$^{3+}$-self-doped rutile TiO$_2$ with visible-light photoactivity, Inorg. Chem. 52 (2013) 3884–3890. doi:10.1021/ic3026182.

[45] S.T. Myung, M. Kikuchi, C.S. Yoon, H. Yashiro, S.J. Kim, Y.K. Sun, B. Scrosati, Black anatase titania enabling ultra high cycling rates for rechargeable lithium batteries, Energy Environ. Sci. 6 (2013) 2609–2614. doi:10.1039/c3ee41960f.

[46] S.G. Ullattil, S.B. Narendranath, S.C. Pillai, P. Periyat, Black TiO$_2$ Nanomaterials: A Review of Recent Advances, Chem. Eng. J. 343 (2018) 708–736. doi:10.1016/j.cej.2018.01.069.

[47] M. Wajid Shah, Y. Zhu, X. Fan, J. Zhao, Y. Li, S. Asim, C. Wang, Facile Synthesis of Defective $TiO_{2-x}$ Nanocrystals with High Surface Area and Tailoring Bandgap for Visible-light Photocatalysis, Sci. Rep. 5 (2015) 1–8. doi:10.1038/srep15804.

[48] X. Chen, D. Zhao, K. Liu, C. Wang, L. Liu, B. Li, Z. Zhang, D. Shen, Laser-Modified Black Titanium Oxide Nanospheres and Their Photocatalytic Activities under Visible Light, ACS Appl. Mater. Interfaces. 7 (2015) 16070–16077. doi:10.1021/acsami.5b04568.

[49] K. Li, J. Xu, X. Yan, L. Liu, X. Chen, Y. Luo, J. He, D.Z. Shen, The origin of the strong microwave absorption in black $TiO_2$, Appl. Phys. Lett. 108 (2016). doi:10.1063/1.4948456.

[50] T. Nakajima, T. Nakamura, K. Shinoda, T. Tsuchiya, Rapid formation of black titania photoanodes: Pulsed laser-induced oxygen release and enhanced solar water splitting efficiency, J. Mater. Chem. A. 2 (2014) 6762–6771. doi:10.1039/c4ta00557k.

[51] M. Yao, J. Zhao, S. Lv, K. Lu, Preparation and hydrogenation of urchin-like titania using a one-step hydrothermal method, Ceram. Int. 43 (2017) 6925–6931. doi:10.1016/j.ceramint.2017.02.115.

[52] Z. Wang, C. Yang, T. Lin, H. Yin, P. Chen, D. Wan, F. Xu, F. Huang, J. Lin, X. Xie, M. Jiang, H-doped black titania with very high solar absorption and excellent photocatalysis enhanced by localized surface plasmon resonance, Adv. Funct. Mater. 23 (2013) 5444–5450. doi:10.1002/adfm.201300486.

[53] F. Teng, M. Li, C. Gao, G. Zhang, P. Zhang, Y. Wang, L. Chen, E. Xie, Preparation of black $TiO_2$ by hydrogen plasma assisted chemical vapor deposition and its photocatalytic activity, Appl. Catal. B Environ. 148–149 (2014) 339–343. doi:10.1016/j.apcatb.2013.11.015.

[54] Y. Yan, B. Hao, D. Wang, G. Chen, E. Markweg, A. Albrecht, P. Schaaf, Understanding the fast lithium storage performance of hydrogenated $TiO_2$ nanoparticles, J. Mater. Chem. A. 1 (2013) 14507–14513. doi:10.1039/c3ta13491a.

[55] Y. Yan, M. Han, A. Konkin, T. Koppe, D. Wang, T. Andreu, G. Chen, U. Vetter, J.R. Morante, P. Schaaf, Slightly hydrogenated $TiO_2$ with enhanced photocatalytic performance, J. Mater. Chem. A. 2 (2014) 12708–12716. doi:10.1039/c4ta02192d.

[56] G. Panomsuwan, A. Watthanaphanit, T. Ishizaki, N. Saito, Water-plasma-assisted synthesis of black titania spheres with efficient visible-light photocatalytic activity, Phys. Chem. Chem. Phys. 17 (2015) 13794–13799. doi:10.1039/C5CP00171D.

[57] Y. Ishida, W. Doshin, H. Tsukamoto, T. Yonezawa, Black $TiO_2$ Nanoparticles by a Microwave-induced Plasma over Titanium Complex Aqueous Solution, Chem. Lett. 44 (2015) 1327–1329. doi:10.1246/cl.150531.

[58] T. Nakano, R. Ito, S. Kogoshi, N. Katayama, Optimal levels of oxygen deficiency in the visible light photocatalyst $TiO_{2-x}$ and long-term stability of catalytic performance, J. Phys. Chem. Solids. (2016). doi:10.1016/j.jpcs.2016.06.017.

[59] G.S. Kim, J.K. Kim, S.H. Kim, J. Jo, C. Shin, J.H. Park, K.C. Saraswat, H.Y. Yu, Specific contact resistivity reduction through Ar plasma-treated $TiO_{2-x}$ interfacial layer to metal/Ge contact, IEEE Electron Device Lett. (2014). doi:10.1109/LED.2014.2354679.

[60] G. Zhu, Y. Shan, T. Lin, W. Zhao, J. Xu, Z. Tian, H. Zhang, C. Zheng, F. Huang, Hydrogenated blue titania with high solar absorption and greatly improved photocatalysis,

Nanoscale. 8 (2016) 4705–4712. doi:10.1039/c5nr07953e.

[61] T. Lin, C. Yang, Z. Wang, H. Yin, X. Lü, F. Huang, J. Lin, X. Xie, M. Jiang, Effective nonmetal incorporation in black titania with enhanced solar energy utilization, Energy Environ. Sci. 7 (2014) 967–972. doi:10.1039/c3ee42708k.

[62] F. Teng, M. Li, C. Gao, G. Zhang, P. Zhang, Y. Wang, L. Chen, E. Xie, Preparation of black $TiO_2$ by hydrogen plasma assisted chemical vapor deposition and its photocatalytic activity, Appl. Catal. B Environ. 148–149 (2014) 339–343. doi:10.1016/j.apcatb.2013.11.015.

[63] H. Wu, C. Xu, J. Xu, L. Lu, Z. Fan, X. Chen, Y. Song, D. Li, Enhanced supercapacitance in anodic $TiO_2$ nanotube films by hydrogen plasma treatment, Nanotechnology. 24 (2013). doi:10.1088/0957-4484/24/45/455401.

[64] B. Bharti, S. Kumar, H.N. Lee, R. Kumar, Formation of oxygen vacancies and $Ti^{3+}$ state in $TiO_2$ thin film and enhanced optical properties by air plasma treatment, Sci. Rep. (2016). doi:10.1038/srep32355.

[65] J. Xu, W. Dong, C. Song, Y. Tang, W. Zhao, Z. Hong, F. Huang, Black rutile (Sn, Ti)$O_2$ initializing electrochemically reversible Sn nanodots embedded in amorphous lithiated titania matrix for efficient lithium storage, J. Mater. Chem. A. 4 (2016) 15698–15704. doi:10.1039/c6ta05645h.

[66] A.P. Singh, N. Kodan, B.R. Mehta, A. Dey, S. Krishnamurthy, In-situ plasma hydrogenated $TiO_2$ thin films for enhanced photoelectrochemical properties, Mater. Res. Bull. (2016). doi:10.1016/j.materresbull.2015.12.015.

[67] Z. Tian, H. Cui, G. Zhu, W. Zhao, J.J. Xu, F. Shao, J. He, F. Huang, Hydrogen plasma reduced black $TiO_2$-B nanowires for enhanced photoelectrochemical water-splitting, J. Power Sources. 325 (2016) 697–705. doi:10.1016/j.jpowsour.2016.06.074.

[68] T. Jedsukontorn, T. Ueno, N. Saito, M. Hunsom, Facile preparation of defective black $TiO_2$ through the solution plasma process: Effect of parametric changes for plasma discharge on its structural and optical properties, J. Alloys Compd. 726 (2017) 567–577. doi:10.1016/j.jallcom.2017.08.028.

[69] S.Z. Islam, A. Reed, S. Nagpure, N. Wanninayake, J.F. Browning, J. Strzalka, D.Y. Kim, S.E. Rankin, Hydrogen incorporation by plasma treatment gives mesoporous black $TiO_2$ thin films with visible photoelectrochemical water oxidation activity, Microporous Mesoporous Mater. 261 (2018) 35–43. doi:10.1016/j.micromeso.2017.10.036.

[70] A. Lepcha, C. Maccato, A. Mettenboerger, T. Andreu, L. Mayrhofer, M. Walter, S. Olthof, T.-P. Ruoko, A. Klein, M. Moseler, K. Meerholz, J.R. Morante, D. Barreca, S. Mathur, Electrospun Black Titania Nanofibres: Influence of Hydrogen Plasma Induced Disorder on the Electronic Structure and Photoelectrochemical Performance, J. Phys. Chem. C. 119 (2015) 18835−18842. doi:10.1021/acs.jpcc.5b02767.

[71] J. xiang Han, Y. lin Cheng, W. bin Tu, T.Y. Zhan, Y. liang Cheng, The black and white coatings on Ti-6Al-4V alloy or pure titanium by plasma electrolytic oxidation in concentrated silicate electrolyte, Appl. Surf. Sci. 428 (2018) 684–697. doi:10.1016/j.apsusc.2017.09.109.

[72] M. Pylnev, W.H. Chang, M.S. Wong, Shell of black titania prepared by sputtering $TiO_2$ target in $H_2$ + Ar plasma, Appl. Surf. Sci. 462 (2018) 285–290.

[73] H.R. An, Y.C. Hong, H. Kim, J.Y. Huh, E.C. Park, S.Y. Park, Y. Jeong, J.I. Park, J.P. Kim, Y.C. Lee, W.K. Hong, Y.K. Oh, Y.J. Kim, M.H. Yang, H.U. Lee, Studies on mass production and highly solar light photocatalytic properties of gray hydrogenated-$TiO_2$ sphere photocatalysts, J. Hazard. Mater. 358 (2018) 222–233. doi:10.1016/j.jhazmat.2018.06.055.

[74] J. Liang, N. Wang, Q. Zhang, B. Liu, X. Kong, C. Wei, D. Zhang, B. Yan, Y. Zhao, X. Zhang, Exploring the mechanism of a pure and amorphous black-blue $TiO_2$:H thin film as a photoanode in water splitting, Nano Energy. 42 (2017) 151–156. doi:10.1016/j.nanoen.2017.10.062.

[75] X. Chen, L. Liu, P.Y. Yu, S.S. Mao, Increasing solar absorption for photocatalysis with black hydrogenated titanium dioxide nanocrystals, Science (80-. ). 331 (2011) 746–750. doi:10.1126/science.1200448.

[76] T. Xia, X. Chen, Revealing the structural properties of hydrogenated black TiO2 nanocrystals, J. Mater. Chem. A. 1 (2013) 2983–2989. doi:10.1039/c3ta01589k.

[77] Z. Wang, C. Yang, T. Lin, H. Yin, P. Chen, D. Wan, F. Xu, F. Huang, J. Lin, X. Xie, M. Jiang, Visible-light photocatalytic, solar thermal and photoelectrochemical properties of aluminium-reduced black titania, Energy Environ. Sci. 6 (2013) 3007–3014. doi:10.1039/c3ee41817k.

[78] J. Cai, Y. Wang, Y. Zhu, M. Wu, H. Zhang, X. Li, Z. Jiang, M. Meng, In Situ Formation of Disorder-Engineered $TiO_2$(B)-Anatase Heterophase Junction for Enhanced Photocatalytic Hydrogen Evolution, ACS Appl. Mater. Interfaces. 7 (2015) 24987–24992. doi:10.1021/acsami.5b07318.

[79] J.Y. Shin, J.H. Joo, D. Samuelis, J. Maier, Oxygen-deficient $TiO_{2-\delta}$ nanoparticles via hydrogen reduction for high rate capability lithium batteries, Chem. Mater. 24 (2012) 543–551. doi:10.1021/cm2031009.

[80] W. Wei, N. Yaru, L. Chunhua, X. Zhongzi, Hydrogenation of $TiO_2$ nanosheets with exposed {001} facets for enhanced photocatalytc activity, RSC Adv. 2 (2012) 8286–8288. doi:10.1039/c2ra21049e.

[81] Y. Yang, J. Liao, Y. Li, X. Cao, N. Li, C. Wang, S. Lin, Electrochemically self-doped hierarchical $TiO_2$ nanotube arrays for enhanced visible-light photoelectrochemical performance: An experimental and computational study, RSC Adv. (2016). doi:10.1039/c6ra05805a.

[82] T. Su, Y. Yang, Y. Na, R. Fan, L. Li, L. Wei, B. Yang, W. Cao, An insight into the role of oxygen vacancy in hydrogenated $TiO_2$ nanocrystals in the performance of dye-sensitized solar cells, ACS Appl. Mater. Interfaces. 7 (2015) 3754–3763. doi:10.1021/am5085447.

[83] X. Lu, G. Wang, T. Zhai, M. Yu, J. Gan, Y. Tong, Y. Li, Hydrogenated $TiO_2$ nanotube arrays for supercapacitors, Nano Lett. (2012). doi:10.1021/nl300173j.

[84] S.S. Mao, X.B. Chen, L. Liu, P.Y. Yu, Increasing Solar Absorption for Photocatalysis with Black Hydrogenated Titanium Dioxide Nanocrystals, Science (80-. ). 331 (2011) 746–750. doi:10.1126/science.1200448.

[85] S.G. Ullattil, P. Periyat, Green microwave switching from oxygen rich yellow anatase to oxygen vacancy rich black anatase $TiO_2$ solar photocatalyst using Mn(II) as "anatase

phase purifier," Nanoscale. (2015). doi:10.1039/c5nr05975e.

[86] X. Xin, T. Xu, J. Yin, L. Wang, C. Wang, Management on the location and concentration of $Ti^{3+}$ in anatase $TiO_2$ for defects-induced visible-light photocatalysis, Appl. Catal. B Environ. (2015). doi:10.1016/j.apcatb.2015.04.016.

[87] G. Zhu, T. Lin, X. Lü, W. Zhao, C. Yang, Z. Wang, H. Yin, Z. Liu, F. Huang, J. Lin, Black brookite titania with high solar absorption and excellent photocatalytic performance, J. Mater. Chem. A. 1 (2013) 9650. doi:10.1039/c3ta11782k.

[88] L. Li, Y. Chen, S. Jiao, Z. Fang, X. Liu, Y. Xu, G. Pang, S. Feng, Synthesis, microstructure, and properties of black anatase and B phase $TiO_2$ nanoparticles, Mater. Des. 100 (2016) 235–240. doi:10.1016/j.matdes.2016.03.113.

[89] H. Wang, T. Lin, G. Zhu, H. Yin, X. Lü, Y. Li, F. Huang, Colored titania nanocrystals and excellent photocatalysis for water cleaning, Catal. Commun. 60 (2015) 55–59. doi:10.1016/j.catcom.2014.11.004.

[90] E.M. Samsudin, S.B.A. Hamid, J.C. Juan, W.J. Basirun, A.E. Kandjani, Surface modification of mixed-phase hydrogenated $TiO_2$ and corresponding photocatalytic response, Appl. Surf. Sci. (2015). doi:10.1016/j.apsusc.2015.10.194.

[91] C. Fan, C. Chen, J. Wang, X. Fu, Z. Ren, G. Qian, Z. Wang, Black Hydroxylated Titanium Dioxide Prepared via Ultrasonication with Enhanced Photocatalytic Activity, Sci. Rep. 5 (2015) 11712. doi:10.1038/srep11712.

[92] M. Ge, J. Cai, J. Iocozzia, C. Cao, J. Huang, X. Zhang, J. Shen, S. Wang, S. Zhang, K.-Q. Zhang, Y. Lai, Z. Lin, A review of $TiO_2$ nanostructured catalysts for sustainable $H_2$ generation, Int. J. Hydrogen Energy. (2017). doi:10.1016/j.ijhydene.2016.12.052.

[93] M. Ge, Q. Li, C. Cao, J. Huang, S. Li, S. Zhang, Z. Chen, K. Zhang, S.S. Al-Deyab, Y. Lai, One-dimensional $TiO_2$ Nanotube Photocatalysts for Solar Water Splitting, Adv. Sci. (2017). doi:10.1002/advs.201600152.

[94] J. Chen, Z. Ding, C. Wang, H. Hou, Y. Zhang, C. Wang, G. Zou, X. Ji, Black Anatase Titania with Ultrafast Sodium-Storage Performances Stimulated by Oxygen Vacancies, ACS Appl. Mater. Interfaces. 8 (2016) 9142–9151. doi:10.1021/acsami.6b01183.

[95] Y.J. He, J.F. Peng, W. Chu, Y.Z. Li, D.G. Tong, Black mesoporous anatase $TiO_2$ nanoleaves: A high capacity and high rate anode for aqueous Al-ion batteries, J. Mater. Chem. A. (2014). doi:10.1039/c3ta13906a.

[96] G. Zhu, J. Xu, W. Zhao, F. Huang, Constructing black titania with unique nanocage structure for solar desalination, ACS Appl. Mater. Interfaces. (2016). doi:10.1021/acsami.6b11466.

[97] S.G. Ullattil, S.B. Narendranath, S.C. Pillai, P. Periyat, Black $TiO_2$ Nanomaterials: A Review of Recent Advances, Chem. Eng. J. 343 (2018) 708–736. doi:10.1016/j.cej.2018.01.069.

[98] C. Zhang, Y. Xie, J. Ma, J. Hu, C. Zhang, A composite catalyst of reduced black $TiO_{2-x}$/CNT: A highly efficient counter electrode for ZnO-based dye-sensitized solar cells, Chem. Commun. 51 (2015) 17459–17462. doi:10.1039/c5cc07284k.

[99] S.G. Ullattil, A.V. Thelappurath, S.N. Tadka, J. Kavil, B.K. Vijayan, P. Periyat, A Sol-solvothermal Processed 'Black $TiO_2$' as Photoanode Material in Dye Sensitized Solar

Cells, Sol. Energy. 155 (2017) 490–495. doi:10.1016/j.solener.2017.06.059.

[100] H. Zhou, Y. Zhang, Electrochemically self-doped $TiO_2$ nanotube arrays for supercapacitors, J. Phys. Chem. C. (2014). doi:10.1021/jp4082883.

[101] C. Kim, S. Kim, J. Lee, J. Kim, J. Yoon, Capacitive and oxidant generating properties of black-colored $TiO_2$ nanotube array fabricated by electrochemical self-doping, ACS Appl. Mater. Interfaces. (2015). doi:10.1021/acsami.5b00123.

[102] J. Zhi, C. Yang, T. Lin, H. Cui, Z. Wang, H. Zhang, F. Huang, Flexible all solid state supercapacitor with high energy density employing black titania nanoparticles as a conductive agent, Nanoscale. 8 (2016) 4054–4062. doi:10.1039/c5nr08136j.

[103] T. Xia, W. Zhang, W. Li, N.A. Oyler, G. Liu, X. Chen, Hydrogenated surface disorder enhances lithium ion battery performance, Nano Energy. (2013). doi:10.1016/j.nanoen.2013.02.005.

[104] T. Xia, W. Zhang, J. Murowchick, G. Liu, X. Chen, Built-in electric field-assisted surface-amorphized nanocrystals for high-rate lithium-ion battery, Nano Lett. (2013). doi:10.1021/nl402810d.

[105] T. Xia, W. Zhang, Z. Wang, Y. Zhang, X. Song, J. Murowchick, V. Battaglia, G. Liu, X. Chen, Amorphous carbon-coated $TiO_2$ nanocrystals for improved lithium-ion battery and photocatalytic performance, Nano Energy. (2014). doi:10.1016/j.nanoen.2014.03.012.

[106] Z. Lu, C.T. Yip, L. Wang, H. Huang, L. Zhou, Hydrogenated $TiO_2$ nanotube arrays as high-rate anodes for lithium-ion microbatteries, Chempluschem. 77 (2012) 991–1000. doi:10.1002/cplu.201200104.

[107] J. Bae, D.S. Kim, H. Yoo, E. Park, Y.G. Lim, M.S. Park, Y.J. Kim, H. Kim, High-Performance $Si/SiO_x$ Nanosphere Anode Material by Multipurpose Interfacial Engineering with Black $TiO_{2-x}$, ACS Appl. Mater. Interfaces. (2016). doi:10.1021/acsami.5b10707.

[108] J.Y. Eom, S.J. Lim, S.M. Lee, W.H. Ryu, H.S. Kwon, Black titanium oxide nanoarray electrodes for high rate Li-ion microbatteries, J. Mater. Chem. A. (2015). doi:10.1039/c5ta01718a.

[109] W. Ren, Y. Yan, L. Zeng, Z. Shi, A. Gong, P. Schaaf, D. Wang, J. Zhao, B. Zou, H. Yu, G. Chen, E.M.B. Brown, A. Wu, A Near Infrared Light Triggered Hydrogenated Black $TiO_2$ for Cancer Photothermal Therapy, Adv. Healthc. Mater. (2015). doi:10.1002/adhm.201500273.

[110] J. Mou, T. Lin, F. Huang, H. Chen, J. Shi, Black titania-based theranostic nanoplatform for single NIR laser induced dual-modal imaging-guided PTT/PDT, Biomaterials. 84 (2016) 13–24. doi:10.1016/j.biomaterials.2016.01.009.

[111] L. Bárdoš, Radio frequency hollow cathodes for the plasma processing technology, Surf. Coatings Technol. 86–87 (1996) 648–656. doi:10.1016/S0257-8972(96)03056-3.

[112] P.F. Little, The Hollow-Cathode Effect and the Theory of Glow Discharges, Proc. R. Soc. A Math. Phys. Eng. Sci. (1954). doi:10.1098/rspa.1954.0152.

[113] T. Degen, M. Sadki, E. Bron, U. König, G. Nénert, The HighScore suite, Powder Diffr. (2014). doi:10.1017/s0885715614000840.

[114] M. Sreemany, S. Sen, A simple spectrophotometric method for determination of the

optical constants and band gap energy of multiple layer TiO$_2$ thin films, Mater. Chem. Phys. 83 (2004) 169–177. doi:10.1016/j.matchemphys.2003.09.030.

[115] W. Chiappim, G.E. Testoni, R.S. Moraes, R.S. Pessoa, J.C. Sagás, F.D. Origo, L. Vieira, H.S. MacIel, Structural, morphological, and optical properties of TiO$_2$ thin films grown by atomic layer deposition on fluorine doped tin oxide conductive glass, Vacuum. 123 (2016) 91–102. doi:10.1016/j.vacuum.2015.10.019.

[116] R. Weingärtner, J.A. Guerra Torres, O. Erlenbach, G. Gálvez De La Puente, F. De Zela, A. Winnacker, Bandgap engineering of the amorphous wide bandgap semiconductor (SiC)$_{1-x}$(AlN)$_x$ doped with terbium and its optical emission properties, in: Mater. Sci. Eng. B Solid-State Mater. Adv. Technol., 2010. doi:10.1016/j.mseb.2010.03.033.

[117] F.H. Wang, J.C. Chao, H.W. Liu, F.J. Liu, Physical properties of TiO$_2$-doped zinc oxide thin films: Influence of plasma treatment in H$_2$ and/or Ar gas ambient, Vacuum. (2017). doi:10.1016/j.vacuum.2016.11.005.

[118] S.J. Baik, J.H. Jang, C.H. Lee, W.Y. Cho, K.S. Lim, Highly textured and conductive undoped ZnO film using hydrogen post-treatment, Appl. Phys. Lett. (1997). doi:10.1063/1.119218.

[119] R. Katal, M. Salehi, M. Hossein, D. Abadi, S. Masudy-panah, S.L. Ong, J. Hu, Preparation of a New Type of Black TiO$_2$ under a Vacuum Atmosphere for Sunlight Photocatalysis, ACS Appl. Mater. Interfaces. 10 (2018) 35316–35326. doi:10.1021/acsami.8b14680.

[120] X. Song, Y. Hu, M. Zheng, C. Wei, Solvent-free in situ synthesis of g-C$_3$N$_4$/{001}TiO$_2$ composite with enhanced UV- and visible-light photocatalytic activity for NO oxidation, Appl. Catal. B Environ. 182 (2016) 587–597. doi:10.1016/j.apcatb.2015.10.007.

[121] G. Žerjav, M.S. Arshad, P. Djinović, J. Zavašnik, A. Pintar, Electron trapping energy states of TiO$_2$–WO$_3$ composites and their influence on photocatalytic degradation of bisphenol A, Appl. Catal. B Environ. 209 (2017) 273–284. doi:10.1016/j.apcatb.2017.02.059.

[122] A.P. Bhirud, S.D. Sathaye, R.P. Waichal, J.D. Ambekar, C.J. Park, B.B. Kale, In-situ preparation of N-TiO$_2$/graphene nanocomposite and its enhanced photocatalytic hydrogen production by H$_2$S splitting under solar light, Nanoscale. 7 (2015) 5023–5034. doi:10.1039/c4nr06435f.

[123] H. Zhang, Z. Xing, Y. Zhang, Z. Li, X. Wu, C. Liu, Q. Zhu, W. Zhou, Ni$^{2+}$ and Ti$^{3+}$ co-doped porous black anatase TiO$_2$ with unprecedented-high visible-light-driven photocatalytic degradation performance, RSC Adv. 5 (2015) 107150–107157. doi:10.1039/c5ra23743b.

[124] H.R. An, Y.C. Hong, H. Kim, J.Y. Huh, E.C. Park, S.Y. Park, Y. Jeong, J.I. Park, J.P. Kim, Y.C. Lee, W.K. Hong, Y.K. Oh, Y.J. Kim, M.H. Yang, H.U. Lee, Studies on mass production and highly solar light photocatalytic properties of gray hydrogenated-TiO$_2$ sphere photocatalysts, J. Hazard. Mater. 358 (2018) 222–233. doi:10.1016/j.jhazmat.2018.06.055.

[125] O. Frank, M. Zukalova, B. Laskova, J. Kürti, J. Koltai, L. Kavan, Raman spectra of titanium dioxide (anatase, rutile) with identified oxygen isotopes (16, 17, 18), Phys. Chem. Chem. Phys. 14 (2012) 14567–14572. doi:10.1039/c2cp42763j.

[126] J. Xu, Z. Tian, G. Yin, T. Lin, F. Huang, Controllable reduced black titania with enhanced photoelectrochemical water splitting performance, Dalt. Trans. 46 (2017) 1047–1051. doi:10.1039/C6DT04060H.

[127] W.F. Zhang, Y.L. He, M.S. Zhang, Z. Yin, Q. Chen, Raman scattering study on anatase $TiO_2$ nanocrystals, J. Phys. D. Appl. Phys. (2000). doi:10.1088/0022-3727/33/8/305.

[128] H. Song, C. Li, Z. Lou, Z. Ye, L. Zhu, Effective Formation of Oxygen Vacancies in Black $TiO_2$ Nanostructures with Efficient Solar-Driven Water Splitting, ACS Sustain. Chem. Eng. 5 (2017) 8982–8987. doi:10.1021/acssuschemeng.7b01774.

[129] L. Wang, D. Wu, Z. Guo, J. Yan, Y. Hu, Z. Chang, Q. Yuan, H. Ming, J. Wang, Ultra-thin $TiO_2$ sheets with rich surface disorders for enhanced photocatalytic performance under simulated sunlight, J. Alloys Compd. 745 (2018) 26–32. doi:10.1016/j.jallcom.2018.02.070.

[130] S. Chen, Y. Wang, J. Li, Z. Hu, H. Zhao, W. Xie, Z. Wei, Synthesis of black $TiO_2$ with efficient visible-light photocatalytic activity by ultraviolet light irradiation and low temperature annealing, Mater. Res. Bull. 98 (2018) 280–287. doi:10.1016/j.materresbull.2017.10.036.

[131] T. Wang, W. Li, D. Xu, X. Wu, L. Cao, J. Meng, A novel and facile synthesis of black $TiO_2$ with improved visible-light photocatalytic $H_2$ generation: Impact of surface modification with CTAB on morphology, structure and property, Appl. Surf. Sci. 426 (2017) 325–332. doi:10.1016/j.apsusc.2017.07.153.

[132] Z. Chen, Y. Zhao, J. Ma, C. Liu, Y. Ma, Detailed XPS analysis and anomalous variation of chemical state for Mn- and V-doped $TiO_2$ coated on magnetic particles, Ceram. Int. 43 (2017) 16763–16772. doi:10.1016/j.ceramint.2017.09.071.

[133] W.S. Tung, W.A. Daoud, New approach toward nanosized ferrous ferric oxide and $Fe_3O_4$-doped titanium dioxide photocatalysts, ACS Appl. Mater. Interfaces. 1 (2009) 2453–2461. doi:10.1021/am900418q.

[134] Y. Sharma, P. Srivastava, A. Dashora, L. Vadkhiya, M.K. Bhayani, R. Jain, A.R. Jani, B.L. Ahuja, Electronic structure, optical properties and Compton profiles of $Bi_2S_3$ and $Bi_2Se_3$, Solid State Sci. 14 (2012) 241–249. doi:10.1016/j.solidstatesciences.2011.11.025.

[135] M. Hannula, H. Ali-Löytty, K. Lahtonen, E. Sarlin, J. Saari, M. Valden, Improved Stability of Atomic Layer Deposited Amorphous $TiO_2$ Photoelectrode Coatings by Thermally Induced Oxygen Defects, Chem. Mater. (2018). doi:10.1021/acs.chemmater.7b02938.

[136] B. Bharti, S. Kumar, H.N. Lee, R. Kumar, Formation of oxygen vacancies and $Ti^{3+}$ state in $TiO_2$ thin film and enhanced optical properties by air plasma treatment, Sci. Rep. (2016). doi:10.1038/srep32355.

[137] X. Chen, L. Liu, Z. Liu, M.A. Marcus, W.C. Wang, N.A. Oyler, M.E. Grass, B. Mao, P.A. Glans, P.Y. Yu, J. Guo, S.S. Mao, Properties of disorder-engineered black titanium dioxide nanoparticles through hydrogenation, Sci. Rep. 3 (2013). doi:10.1038/srep01510.

[138] J. Huo, Y. Hu, H. Jiang, C. Li, In situ surface hydrogenation synthesis of $Ti^{3+}$ self-doped $TiO_2$ with enhanced visible light photoactivity, Nanoscale. (2014). doi:10.1039/c4nr00972j.

[139] H. Albetran, B.H. O'Connor, I.M. Low, Effect of calcination on band gaps for electrospun

titania nanofibers heated in air-argon mixtures, Mater. Des. (2016). doi:10.1016/j.matdes.2015.12.044.

[140] S. Chen, Y. Wang, J. Li, Z. Hu, H. Zhao, W. Xie, Z. Wei, Synthesis of black $TiO_2$ with efficient visible-light photocatalytic activity by ultraviolet light irradiation and low temperature annealing, Mater. Res. Bull. 98 (2018) 280–287. doi:10.1016/j.materresbull.2017.10.036.

[141] D.M. Mattox, Particle bombardment effects on thin-film deposition: A review, J. Vac. Sci. Technol. A Vacuum, Surfaces, Film. (2002). doi:10.1116/1.576238.

[142] R.E. Cuthrell, D.M. Mattox, C.R. Peeples, P.L. Dreike, K.P. Lamppa, Residual stress anisotropy, stress control, and resistivity in post cathode magnetron sputter deposited molybdenum films, J. Vac. Sci. Technol. A Vacuum, Surfaces, Film. (2002). doi:10.1116/1.575451.

[143] S. Wendt, P.T. Sprunger, E. Lira, G.K.H. Madsen, Z. Li, J. Hansen, J. Matthiesen, A. Blekinge-Rasmussen, E. Lægsgaard, B. Hammer, F. Besenbacher, The role of interstitial sites in the Ti$3d$ defect state in the band gap of titania, Science (80-.). (2008). doi:10.1126/science.1159846.

[144] R. Sanjinés, H. Tang, H. Berger, F. Gozzo, G. Margaritondo, F. Lévy, Electronic structure of anatase $TiO_2$ oxide, J. Appl. Phys. (1994). doi:10.1063/1.356190.

[145] C.C. Fulton, G. Lucovsky, R.J. Nemanich, Electronic states at the interface of Ti–Si oxide on Si(100), J. Vac. Sci. Technol. B Microelectron. Nanom. Struct. (2002). doi:10.1116/1.1493785.

[146] A.L.J. Pereira, P.N. Lisboa Filho, J. Acua, I.S. Brandt, A.A. Pasa, A.R. Zanatta, J. Vilcarromero, A. Beltrán, J.H. Dias Da Silva, Enhancement of optical absorption by modulation of the oxygen flow of $TiO_2$ films deposited by reactive sputtering, in: J. Appl. Phys., 2012. doi:10.1063/1.4724334.

[147] M.T. Greiner, Z.H. Lu, Thin-film metal oxides in organic semiconductor devices: Their electronic structures, work functions and interfaces, NPG Asia Mater. (2013). doi:10.1038/am.2013.29.

[148] C. Di Valentin, G. Pacchioni, A. Selloni, Reduced and n-type doped $TiO_2$: Nature of $Ti^{3+}$ species, J. Phys. Chem. C. (2009). doi:10.1021/jp9061797.

[149] Y. Chen, Q. Wu, L. Liu, J. Wang, Y. Song, The fabrication of self-floating $Ti^{3+}$/N co-doped $TiO_2$/diatomite granule catalyst with enhanced photocatalytic performance under visible light irradiation, Appl. Surf. Sci. 467–468 (2019) 514–525. doi:10.1016/j.apsusc.2018.10.146.

[150] Y. Chen, K. Liu, Fabrication of magnetically recyclable Ce/N co-doped $TiO_2$/$NiFe_2O_4$/diatomite ternary hybrid: Improved photocatalytic efficiency under visible light irradiation, J. Alloys Compd. (2017). doi:10.1016/j.jallcom.2016.12.153.

[151] K. Zhang, W. Zhou, X. Zhang, B. Sun, L. Wang, K. Pan, B. Jiang, G. Tian, H. Fu, Self-floating amphiphilic black $TiO_2$ foams with 3D macro-mesoporous architectures as efficient solar-driven photocatalysts, Appl. Catal. B Environ. 206 (2017) 336–343. doi:10.1016/j.apcatb.2017.01.059.

[152] A.A. Ismail, I. Abdelfattah, A. Helal, S.A. Al-Sayari, L. Robben, D.W. Bahnemann, Ease synthesis of mesoporous $WO_3$-$TiO_2$ nanocomposites with enhanced photocatalytic

performance for photodegradation of herbicide imazapyr under visible light and UV illumination, J. Hazard. Mater. (2016). doi:10.1016/j.jhazmat.2015.12.041.

[153] X. Pan, M.Q. Yang, X. Fu, N. Zhang, Y.J. Xu, Defective $TiO_2$ with oxygen vacancies: Synthesis, properties and photocatalytic applications, Nanoscale. (2013). doi:10.1039/c3nr00476g.

[154] Z. Bian, T. Tachikawa, P. Zhang, M. Fujitsuka, T. Majima, A nanocomposite superstructure of metal oxides with effective charge transfer interfaces, Nat. Commun. (2014). doi:10.1038/ncomms4038.

**Table I -** TiO$_2$ thin films deposition parameters.

| Parameter | Value |
|---|---|
| Ar flow (sccm) | 15.0 |
| O$_2$ flow (sccm) | 3.0 |
| Working pressure (Pa) | 1.3 |
| Residual pressure (Pa) | 5.0×10$^{-3}$ |
| Deposition time (min) | 30 |
| DC power (W) | 150 |
| Film thickness (nm) | 500 – 600 |

**Table II -** Hydrogenation plasma process parameters.

| Parameter | Value |
|---|---|
| H$_2$ flow rate (sccm) | 45 |
| Working pressure (Pa) | 13.3 |
| Residual pressure (Pa) | 5.0×10$^{-2}$ |
| RF power (W) | 200 |
| Bias voltage (V) | 380 |
| Treatment time (min) | 0, 15, 30, 45, 60 |

**Table III -** Surface area evaluated by AFM (1 μm × 1 μm scanning area), and sheet resistance.

| Samples | Surface area (μm$^2$) | kΩ/□ |
|---|---|---|
| Pristine TiO$_2$ | 1.03 | >10$^3$ |
| Black 15 min | 1.28 | 1.93 ± 0.05 |
| Black 30 min | 1.12 | 1.40 ± 0.02 |
| Black 45 min | 1.11 | 1.42 ± 0.07 |
| Black 60 min | 1.10 | 0.95 ± 0.14 |

**Table IV -** Binding energy and percentage of integrated areas of the high resolution Ti 2p XPS curves of pristine and black $TiO_2$ films.

| Ti Species | Pristine $TiO_2$ | | Black 15 | | Black 30 | | Black 45 | | Black 60 | |
|---|---|---|---|---|---|---|---|---|---|---|
| | BE (eV) | % Area | BE (eV) | % Area | BE (eV) | % Area | BE (eV) | % Area | BE (eV) | % Area |
| $Ti^{3+}2p_{3/2}$ | 459.69 | 100 | 458.96 | 72.52 | 458.91 | 72.37 | 458.74 | 71.78 | 458.92 | 68.65 |
| $Ti^{2+}2p_{3/2}$ | - | - | 456.93 | 13.84 | 456.93 | 15.06 | 456.76 | 15.78 | 456.93 | 16.44 |
| $Ti^{2+}2p_{3/2}$ | - | - | 455.54 | 13.64 | 455.56 | 12.57 | 455.39 | 12.44 | 455.53 | 14.92 |

**Table V -** Binding energy and percentage of integrated areas of the high-resolution O1s XPS curves of pristine and black $TiO_2$ films.

| Samples | Area O-H Peak (%) | Area Ti-O Peak (%) |
|---|---|---|
| Pristine $TiO_2$ | 25.2 | 74.8 |
| Black 15 | 44.9 | 55.1 |
| Black 30 | 45.1 | 54.9 |
| Black 45 | 47.4 | 52.6 |
| Black 60 | 47.5 | 52.5 |

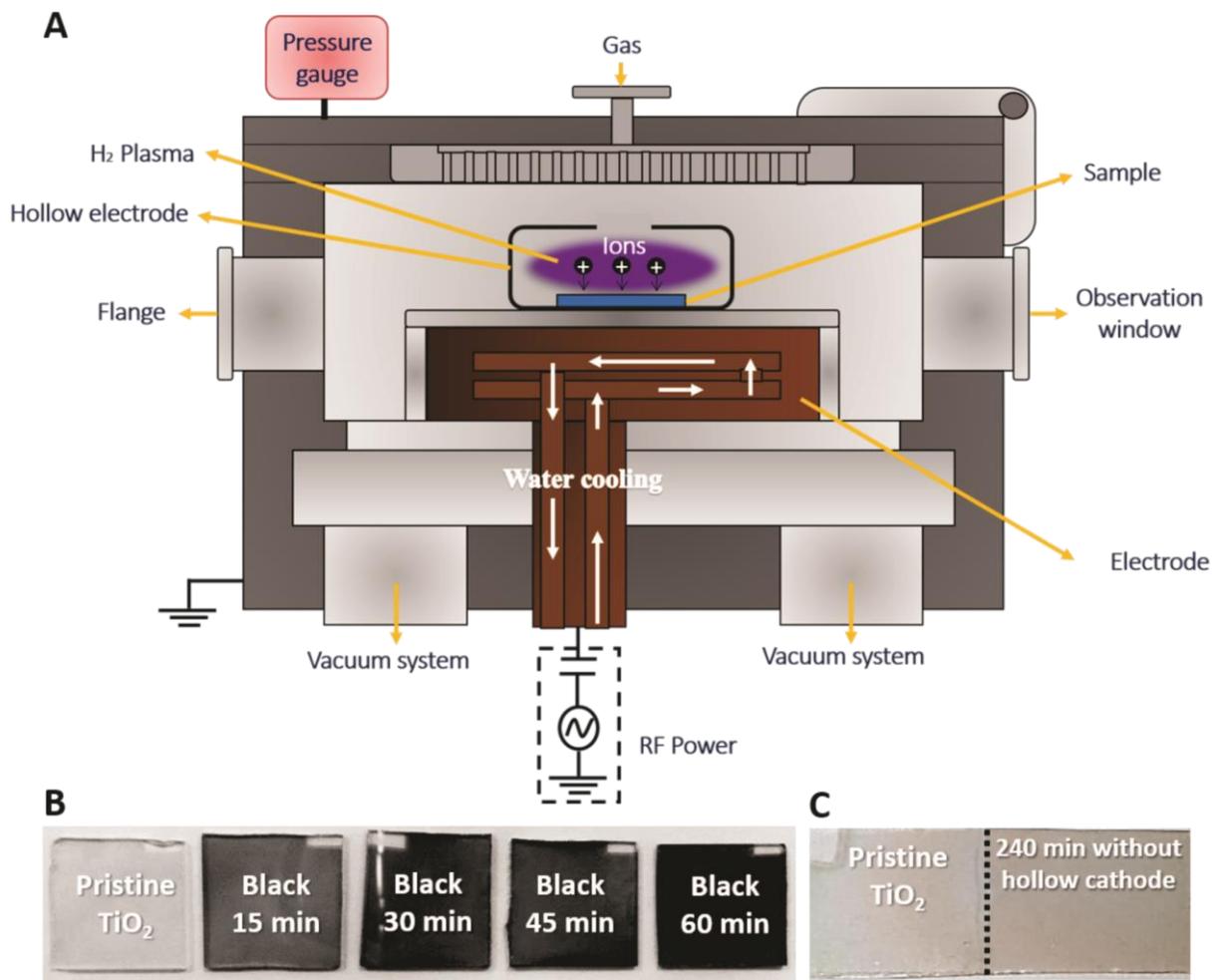

**Figure 1.** (a) Schematic diagram of the hollow cathode H$_2$ plasma reactor used for the TiO$_2$ thin films hydrogenation. Photograph of the TiO$_2$ films before and after the treatment in hydrogen plasma (b) using the hollow catode and (c) without the hollow cathode (planar electrode).

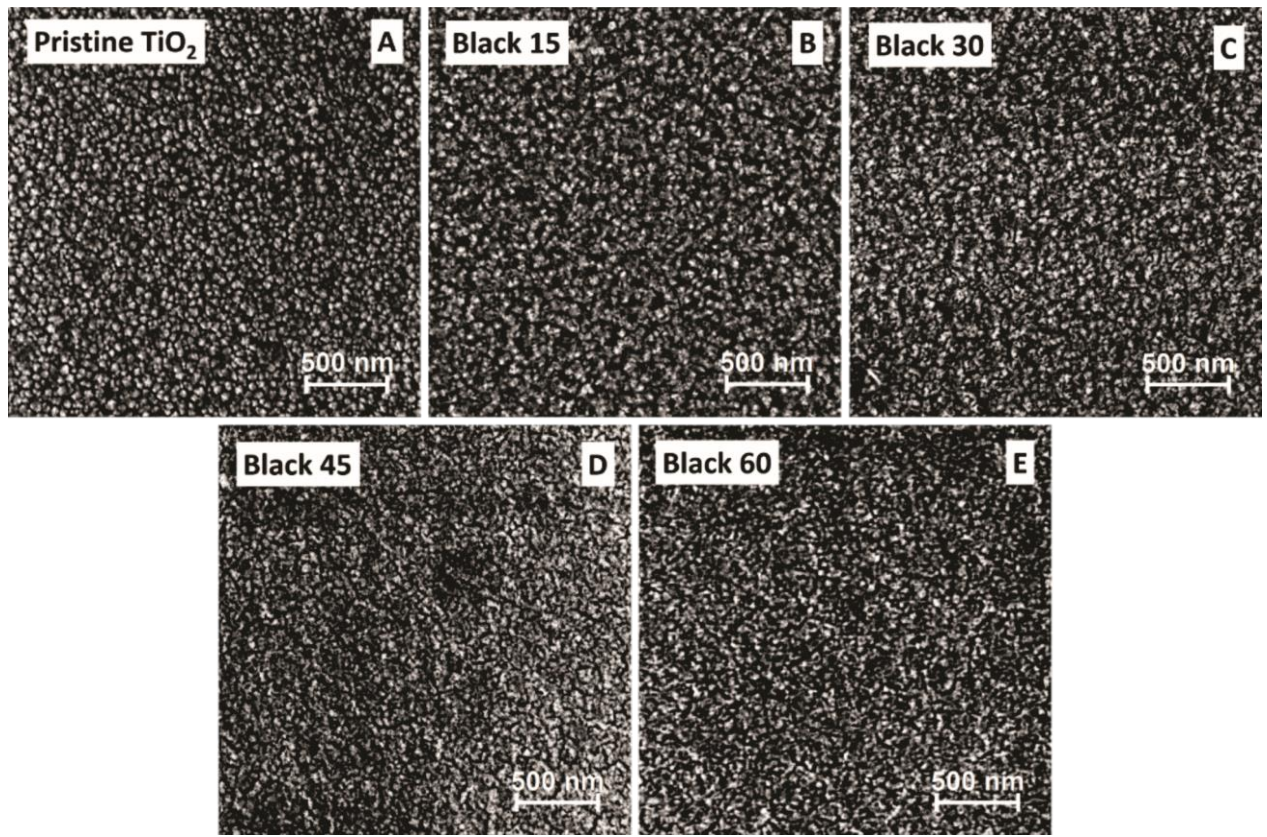

**Figure 2.** FEG-SEM images of the produced $TiO_2$ films deposited on c-Si substrates: (a) as-grown pristine and black $TiO_2$ thin films subjected to hydrogen plasma at different times: (b) 15, (c) 30, (d) 45, and (e) 60 min.

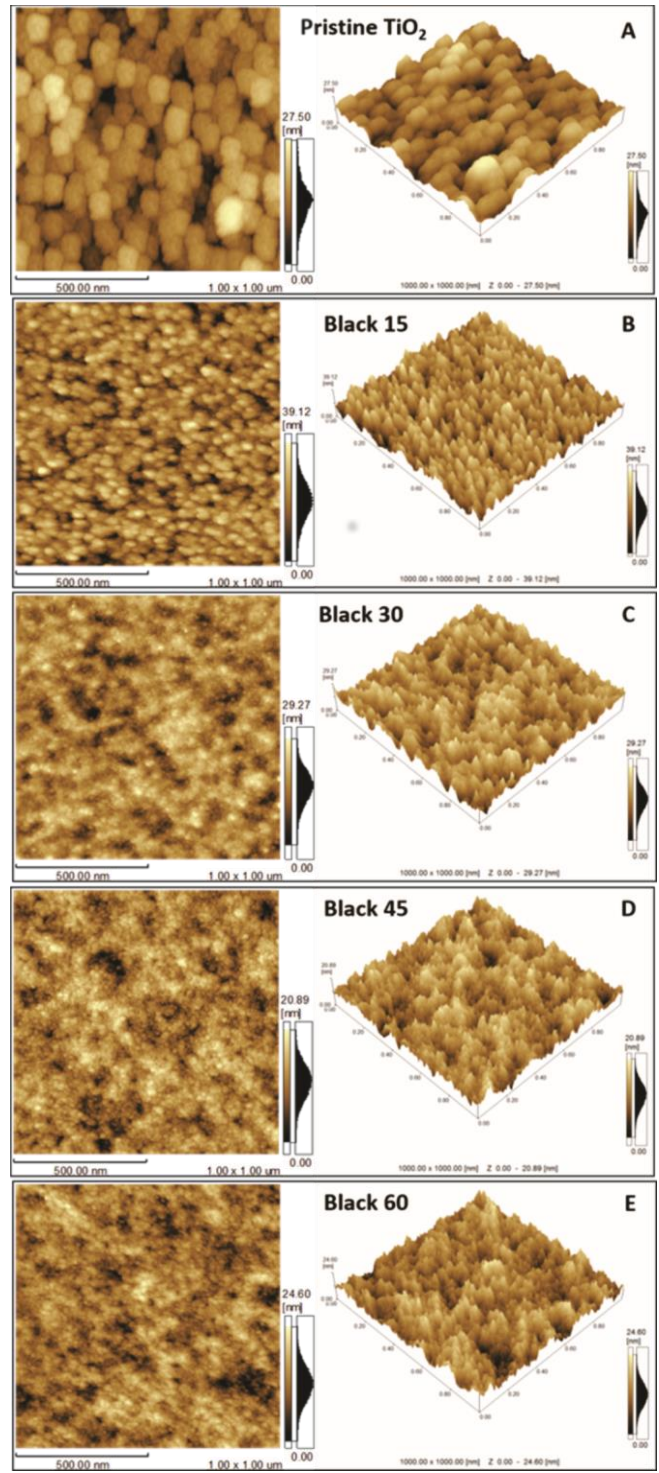

**Figure 3.** AFM images (1 μm × 1 μm scanning area) of the produced TiO$_2$ films deposited on c-Si substrates: (a) as-grown pristine and black TiO$_2$ thin films treated in hydrogen plasma at different times: (b) 15, (c) 30, (d) 45, and (e) 60 min.

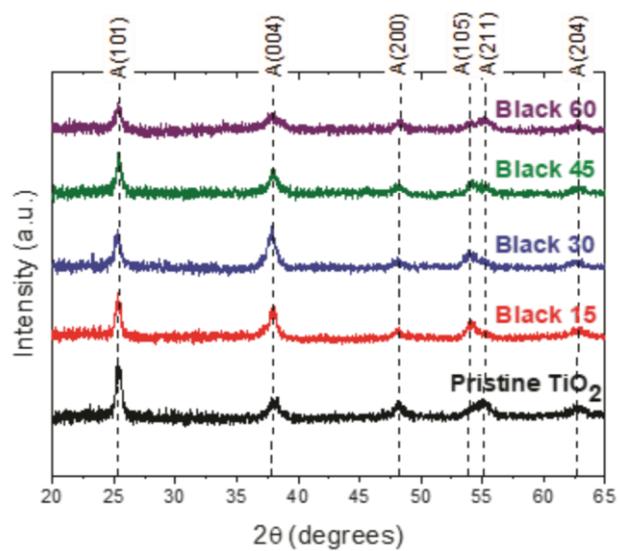

**Figure 4.** XRD spectra of the pristine and black TiO$_2$ samples treated by hydrogen plasma during 15, 30, 45, and 60 min. The main characteristic peaks of anatase TiO$_2$ are indicated.

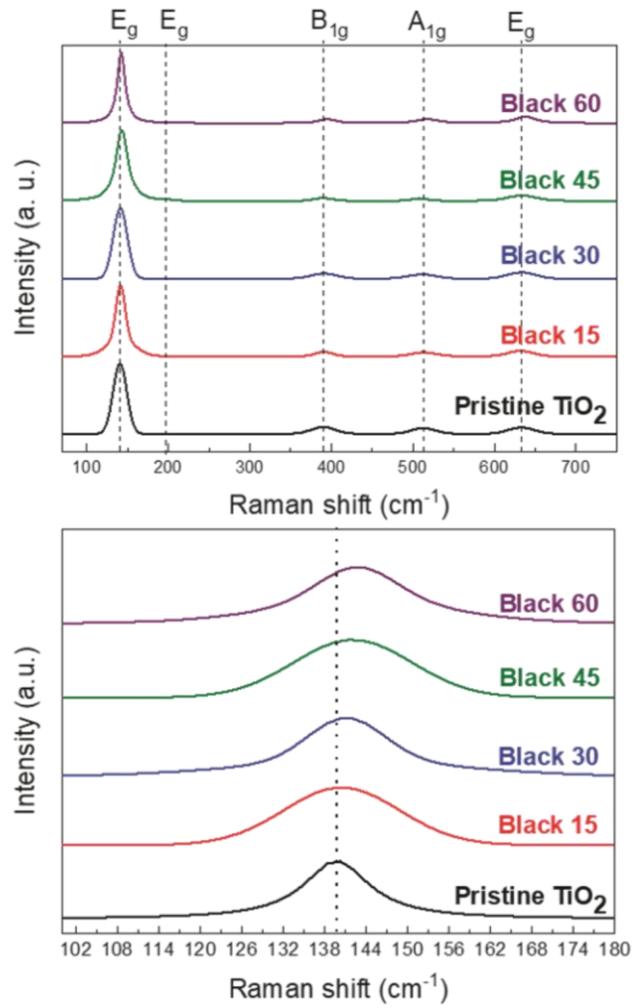

**Figure 5.** (a) Raman spectra of the pristine and black TiO$_2$ samples treated by hydrogen plasma at different times (15, 30, 45, and 60 min) and (b) focused on the anatase E$_g$ mode peak.

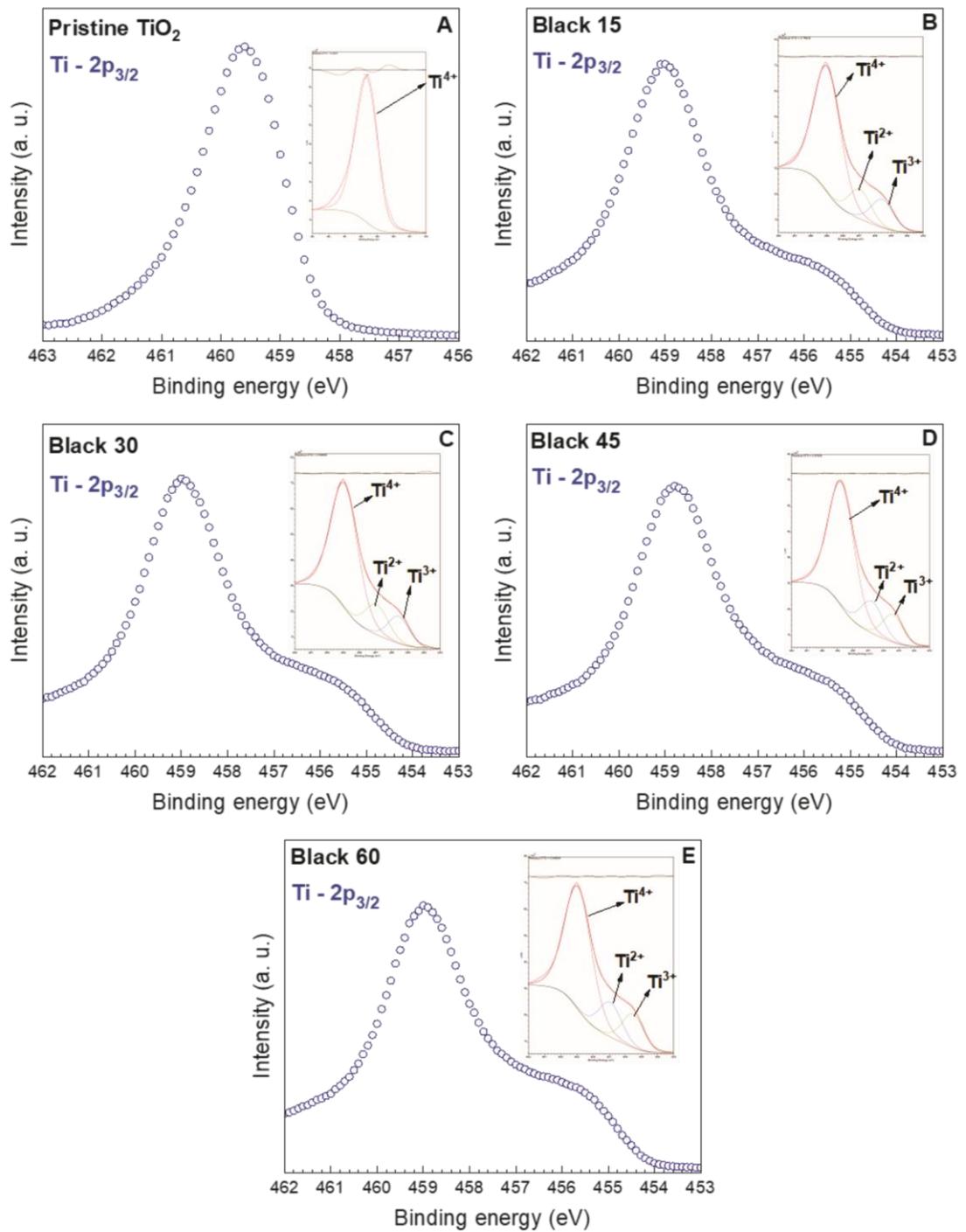

**Figure 6.** High resolution XPS spectra of Ti 2p$_{3/2}$ peaks of (a) pristine and black TiO$_2$ thin films treated in hydrogen plasma during (b) 15, (c) 30, (d) 45, and (e) 60 min.

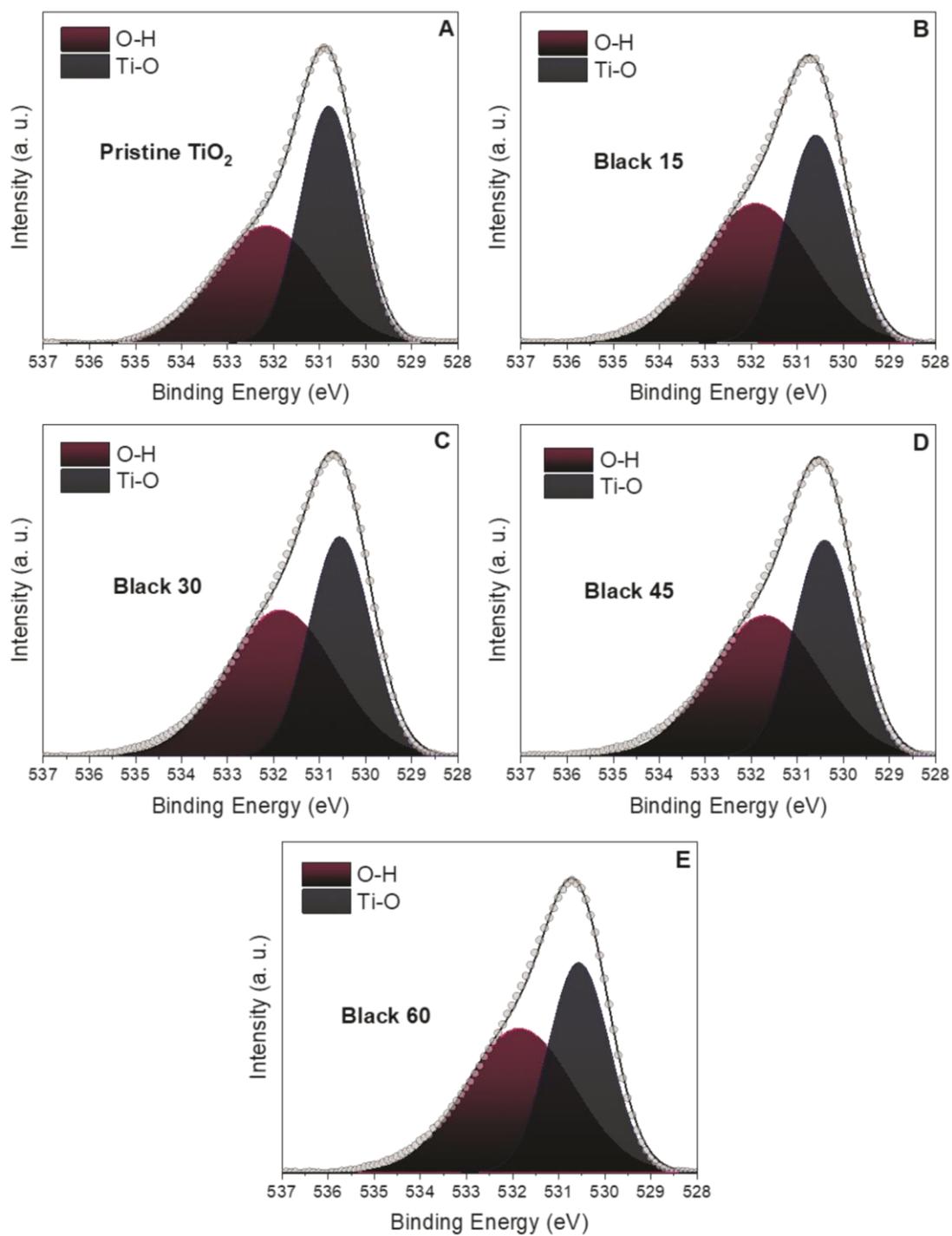

**Figure 7.** High Resolution XPS spectra of O 1s peaks (a) pristine and black TiO$_2$ thin films treated in hydrogen plasma during (b) 15, (c) 30, (d) 45, and (e) 60 min. Open circles and line represents the experimental data and the fit, respectively.

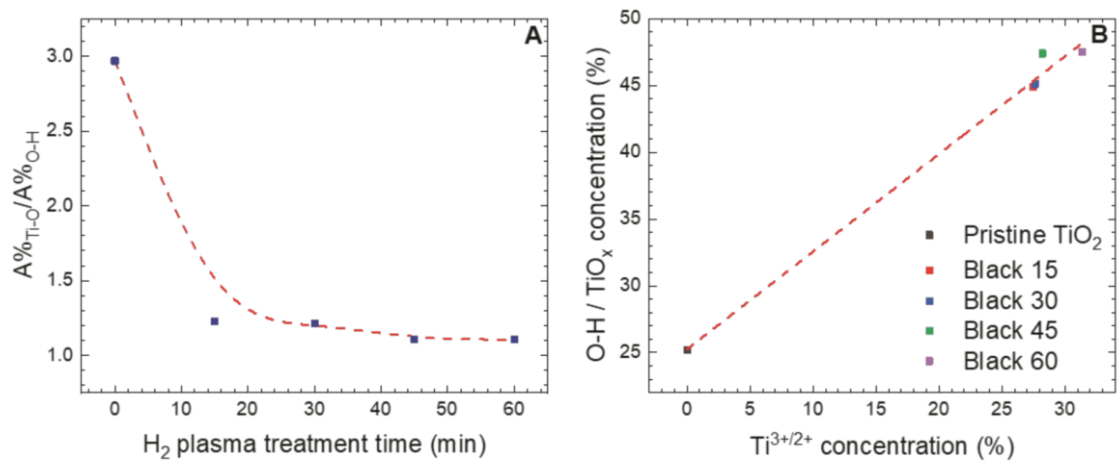

**Figure 8.** (a) Peak percentage area ratio $A\%_{Ti-O}/A\%_{O-H}$ between Ti-O and O-H species as the hydrogen plasma treatment time increase. The dashed red line is just a guide for the eyes. (b) Defective O concentration as function of relative $Ti^{3+/2+}$ suboxides concentration for films treated in hydrogen plasma at different times. The red dashed line is a linear fit considering all the points.

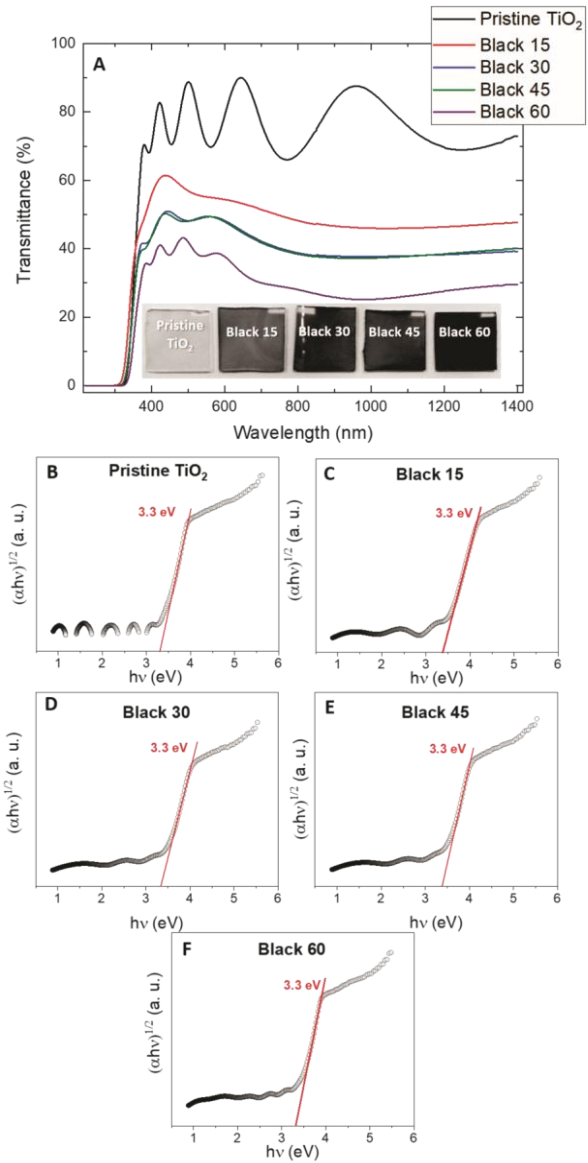

**Figure 9.** (a) UV-vis-NIR transmittance spectra of pristine and plasma treated $TiO_2$ films, Tauc's plot curves of pristine (b) and black $TiO_2$ samples treated by hydrogen plasma at (c) 15, (d) 30, (e) 45, and (f) 60 min.

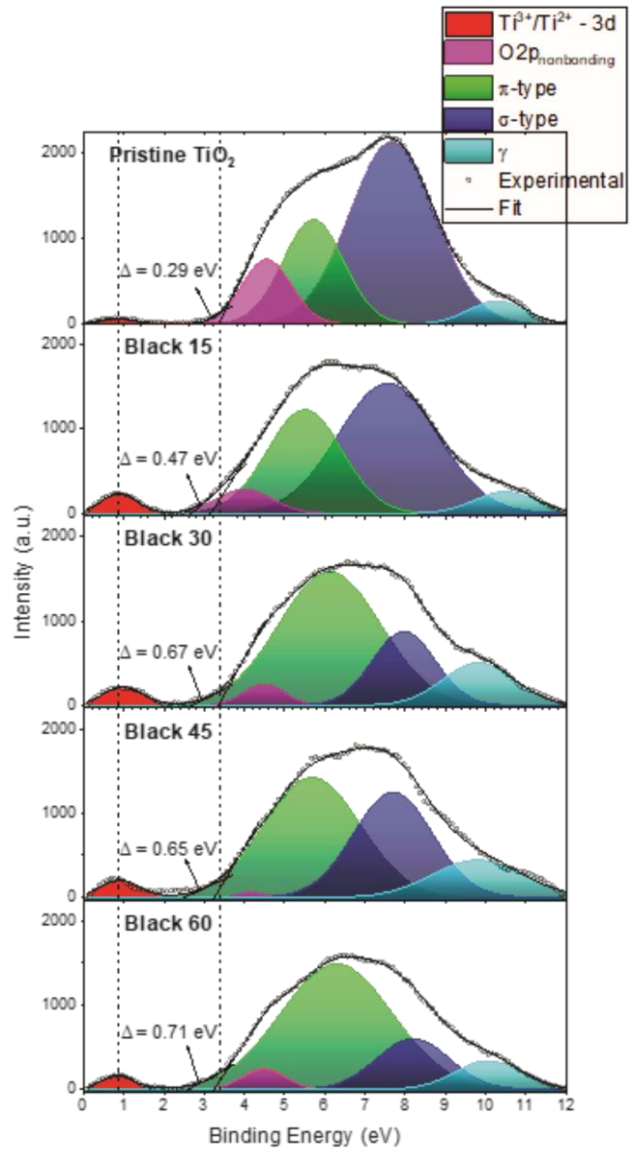

**Figure 10.** XPS valance band spectra of pristine and black TiO$_2$ thin films treated by hydrogen plasma during 15, 30, 45, and 60 min.

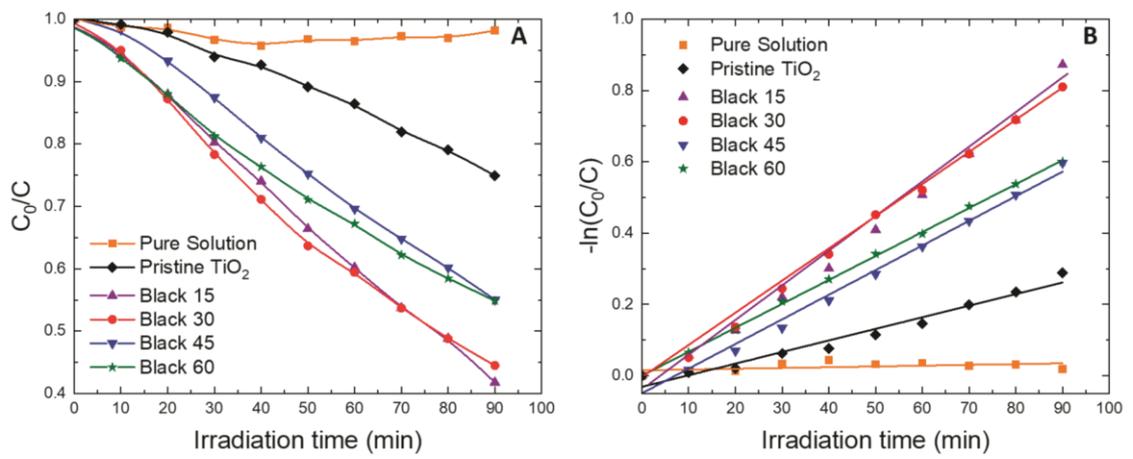

**Figure 11.** (a) Time dependence of methylene blue relative concentration over pristine and black TiO$_2$ thin films under UV light irradiation and (b) the kinect plots for each sample.